# RIGOROUS STATISTICAL DETECTION AND CHARACTERIZATION OF A DEVIATION FROM THE GUTENBERG-RICHTER DISTRIBUTION ABOVE MAGNITUDE 8 IN SUBDUCTION ZONES

## V.F.Pisarenko[1] and D.Sornette[2,3]

[1]International Institute of Earthquake Prediction Theory and Mathematical Geophysics
Russian Ac. Sci. Warshavskoye sh., 79, kor. 2, Moscow 113556, Russia
[2]Institute of Geophysics and Planetary Physics and Department of Earth and Space Science
University of California, Los Angeles, California 90095
[3]Laboratoire de Physique de la Matière Condensée, CNRS UMR6622 and Université des Sciences
B.P. 70, Parc Valrose, 06108 Nice Cedex 2, France

**Abstract**

We present a quantitative statistical test for the presence of a crossover $c_0$ in the Gutenberg-Richter distribution of earthquake seismic moments, separating the usual power law regime for seismic moments less than $c_0$ from another faster decaying regime beyond $c_0$. Our method is based on the transformation of the ordered sample of seismic moments into a series with uniform distribution under condition of no crossover. The bootstrap method allows us to estimate the statistical significance of the null hypothesis $H_0$ of an absence of crossover ($c_0$=infinity). When $H_0$ is rejected, we estimate the crossover $c_0$ using two different competing models for the second regime beyond $c_0$ and the bootstrap method. For the catalog obtained by aggregating 14 subduction zones of the Circum Pacific Seismic Belt, our estimate of the crossover point is $\log(c_0)$ =28.14 ± 0.40 ($c_0$ in dyne-cm), corresponding to a crossover magnitude $m_W$=8.1± 0.3. For separate subduction zones, the corresponding estimates are much more uncertain, so that the null hypothesis of an identical crossover for all subduction zones cannot be rejected. Such a large value of the crossover magnitude makes it difficult to associate it directly with a seismogenic thickness as proposed by many different authors in the past. Our measure of $c_0$ may substantiate the concept that the localization of strong shear deformation could propagate significantly in the lower crust and upper mantle, thus increasing the effective size beyond which one should expect a change of regime.

# 1-Introduction

Earthquakes exhibit considerable complexity in their organization both in space and time but have also strong regularities. The most famous and best-established one is the Gutenberg-Richer (G-R) size-frequency relationship giving the number $N(m_W)$ of earthquakes of magnitude larger than $m_W$ (in a large given geographic area over a long time interval) (Gutenberg and Richter, 1954). Translating the magnitude $m_W$=(2/3)$\log_{10} M_W$-6 in seismic moment $M_W$=g d S expressed in N-m units (where g is an average shear elastic coefficient of the crust, d is the average slip of the earthquake over a surface S of rupture), the Gutenberg-Richter law gives the number $N(M_W)$ of earthquake of seismic moment larger than $M_W$. The striking empirical observation is that $N(M_W)$ can be modeled with a very good approximation by a power law

$$(1) \qquad N(M_W) \sim M_W^{-\beta},$$

where $\beta$=(2/3)b and the b-value is approximately 1 thus giving $\beta \approx 2/3$. The Gutenberg-Richter law (1) is found to hold over a large interval of seismic moments ranging from $10^{20} \div 10^{24}$ ($m_W$=2.6-4) to about $10^{26.5}$ dyne-cm ($m_W$=7). Many works have investigated possible variations of this law (1) from one seismic region to another and as a function of magnitude and time. Two main deviations have been reported and discussed repeatedly in the literature:

1) from general energy considerations, the power law (1) has to cross-over at a "corner" magnitude to a faster decaying law. This would translate into a downward bend in the linear frequency-magnitude log-log plot of (1). The corner magnitude has been estimated to be 7.5 for subduction zones and 6.0 for midocean ridge zones (*Pacheco et al.,*1992), (*Okal and Romanowicz,* 1994) but this is hotly debated (see below);

2) the exponent b is different in subduction and in midocean-ridge zones; there is in addition a controversy among seismologists about the homogeneity of b-values in different zones of the same tectonic type. Some seismologists believe that b-values are different at least in several zone groups, others find these differences statistically insignificant.

 With respect to the first point, a number of authors have argued for a change of the frequency distribution from small to large events based on the idea that small earthquakes and large earthquakes are not self-similar due to the existence of characteristic scales, such as the thickness of the seismogenic crust. Roughly speaking, the finite thickness of the seismogenic crust restricts the accumulation of elastic energy in 3-D volumes, thus slightly discriminating large events, leading to the so-called loss of one dimension by the earthquake source (see *Kanamori and Anderson*, 1975; *Main and Burton*, 1984; *Rundle,* 1989; *Romanowicz*, 1992; *Pacheco et al.,* 1992; *Romanowicz and Rundle,* 1993; *Okal and Romanowicz*, 1994; *Sornette et al*., 1996; *Molchan et al.,* 1996, 1997; *Kagan*, 1997; 1999; *Sornette and Sornette,* 1999). Moreover, the power law distribution that holds for small earthquakes cannot be extended to infinite magnitudes because it would require that an infinite amount of energy be released from the Earth's interior. Thus, it can be concluded with certainty from this energetic argument that the magnitude-frequency law has to eventually bend down in its extreme tail. Therefore, the relevant question is not whether this downward bend occurs but rather whether the magnitude range over which the crossover occurs can be observed and estimated reliably. For instance, there is nothing fundamental that prevents in principle the corner magnitude to be, say, $m_W$=10. Such a value would ensure the finiteness of the earthquake energy flow over long times, but would be unobservable in presently available catalogs.

With respect to the second point, detailed studies of the spatial variability of seismic parameters can be found in (*Kronrod*, 1984; *Cornell*, 1994; *Kagan*, 1997; 1999; *Molchan et al.*1996). The worldwide seismicity is usually studied using the Flinn-Engdahl regionalization or some of its modifications (*Flinn et al*., 1974; *Kronrod*, 1984; *Young et al*., 1996). Using a more coarse-grained regionalization and new statistical tests, Pisarenko and Sornette (2001) confirmed the already documented observation that the slope b of the Gutenberg-Richter law for shallow events is significantly smaller for subduction zones (SZ) compared to mid-ocean ridge zones (MORZ) (*Okal and Romanovicz*, 1994; *Kagan*, 1997, 1999; *Molchan et al.,*1996). Neither a statistical scatter nor a lower seismic flux of MORZ can mask this difference. They propose that the large value b ≈ 1.5 of MORZ earthquakes, (at least for the transform earthquakes constituting the most numerous and powerful fraction of all oceanic events) with the largely extensional stress configuration and the presence of abundant water, result from the fact that faults remain weak and open. In contrast, the smaller value b ≈ 1 found for subduction zones could be interpreted as the signature of fast healing faults with a larger compressional component of stress. In a recent analysis, *Bird et al*. (2000) explain the difference of b-value found earlier (*Okal and Romanovicz, 1994; Kagan,* 1997, 1999; *Molchan et al.,*1996; *Pisarenko and Sornette,* 2001) from the fact that an effective larger b-value will be found when mixing power law distribution with different "corner" magnitudes corresponding to two types of earthquake sources: strike-slip and normal faults. However, as was shown by Pisarenko and Sornette (2001), the separate analysis of MOR-events with different types of source (strike-slip and normal fault) has confirmed the significantly larger b-values for MOR zones.

 The authors  (*Kanamori and Anderson*, 1975; *Main and Burton*, 1984; *Rundle*, 1989; *Romanowicz*, 1994; *Pacheco et al*., 1992; *Pacheco and Sykes*, 1992; *Romanowicz and Rundle*, 1993; *Okal and Romanowicz*, 1994; *Sornette et al*., 1996; *Molchan et al*., 1996, 1997; *Kagan*, 1997; 1999; *Sornette and Sornette*, 1999) propose that the large-magnitude branch of the distribution can be modeled also by a power-like law and that the crossover moment or magnitude between these two distributions can be connected with the thickness of the seismogenic zone. Pacheco et al. (1992) claim to have identified a kink in the distribution of shallow transform fault earthquakes in MOR around magnitude 5.9 to 6.0, which corresponds to a characteristic dimension of about 10 km; a kink for subduction zones is presumed to occur at a moment magnitude near 7.5, which corresponds to a downdip dimension of the order of 60 km. However, Sornette et al. (1996) have shown that this claim cannot be defended convincingly because the crossover magnitude between the two regimes is ill-defined.

Since the largest earthquakes contribute a significant fraction of the total deformation budget of the crust, determining their frequency is of paramount importance for seismic risk assessment. Also, from a physical point of view, the value of the "corner" magnitude and the shape of the frequency-size distribution beyond it may provide insight in the underlying mechanism and constraint the modeling. It is however fair to say that the status on the detection of a change of regime in the Gutenberg-Richter law is still very much open and controversial. Here, we attempt to address the following questions.

(a) Is there indeed a detectable and statistically significant crossover of the Gutenberg-Richter law to a faster decaying law for the very largest observed earthquakes? Does the answer to this question depend on the size of the catalog in question?

(b) What is the uncertainty of the estimation of the "corner" magnitude at which this cross-over occurs, if it exists?

(c) What can be said about the form of the frequency-size distribution beyond the "corner" magnitude?

(d) What conclusions can be obtained about possible variations of b-values?

To address these questions, Pacheco et al. (1992) relied on visual inspection, Sornette et al. (1996) on Monte-Carlo simulations, Kagan (1997, 1999) and Kagan and Schoenberg (2001) on maximum likelihood estimation of a postulated Pareto distribution tapered by an exponential. Several parametric families, such as Gamma distributions (*Main and Burton*, 1984; *Main*, 1996; *Kagan*, 1994, 1997), modified Pareto distribution (*Kagan and Schoenberg*, 2001), two power law distributions with a crossover point (*Sornette et al.* 1996) and Weibull distributions (*Laherrere and Sornette*, 1998) were suggested for earthquake energy distributions including the tail range, but none of these models is universally accepted. A detailed study of this problem leads us to a conservative conclusion (*Pisarenko and Sornette*, 2001): none of the suggested laws is preferable because of a very small number of observations in the extreme range. In other words, these different families of distributions are practically undistinguishable, given the available data.

To our knowledge, there has not been any systematic statistical approach which addresses the questions (a)-(d) independently. In other words, previous attempts have consisted of tests of the joined hypothesis that there is a cross-over at some "corner" magnitude with some assumed functional form. We propose here a novel statistical approach that addresses the questions (a)-(c) sequentially. In this way, we obtain a novel and efficient statistical test of the possible deviations from a power-like law, based on the properties of the order statistics of catalogs. The existence of possible deviations and the value of the "corner" frequency can thus be discussed independently of any assumption of the parametric form of the extreme tail of the magnitude-frequency distribution. We feel that this is a very important step towards resolving unambiguously the issues raised by previous works and the questions (a)-(d).

The organization of this paper is as follows. In the next section, we describe the data, the definition of the tectonic zones and the corresponding catalogs. Section 3 introduces the Generalized Pareto Distribution introduced in the earthquake concept as an improvement over the Gutenberg-Richter law by Dargahi-Noubary (1986), see as well Pisarenko and Sornette (2001), Dargahi-Noubary (2001), and summarizes its main properties. Section 4 describes our novel statistical test for a deviation from the power law behavior (1) in the tail of earthquake size distributions. Section 5 describes the determination of the corner magnitude for catalogs for which the method of section 4 has concluded positively about the existence of a change of regime. Section 6 presents a discussion of our results and concludes.

## 2-Data sets

We used the Harvard catalog of seismic moments covering the period 01.01.1977 to 31.05.2000 (*Dziewonski et al.*, 1994). Since the distribution of earthquake energy for deep events differ significantly from that of the shallow ones (*Kagan*, 1997), we restrict our analysis to shallow earthquakes with focal depth h < 70 km. Such events constitute about 75% of the catalog. In order to illustrate our detailed analysis of seismic moment distributions, we have chosen subduction zones of the Circum Pacific Seismic Belt (CPSB). The main part of the total world seismic energy is radiated in this region.

All zones of the CPSB constitute a group of relatively homogeneous zones from a tectonic viewpoint, whose dynamics is governed by the subduction process. Modern plate tectonics defines 32 subduction zones in the CPSB (see *Jarrard*,1986)). The smallest zones contain 50 to 100 shallow events recorded by the Harvard catalog, in the period 1977-2000. This number is too small for our detailed statistical analysis. Therefore, we had to unite some small zones in order to provide samples of size at least $n \cong 170$, which is a minimum requirement for the statistical technique that we introduce. As a result of this aggregation procedure, we have formed 12 larger subduction zones in the CPSB. Their parameters are given in Table 1. For our analysis, we have added one subduction zone at the boundary of the Indian Ocean (Sunda), that presents a high seismicity.

For a collective analysis of all subduction zones, we have also added one small zone: New Guinea (n=128, M ≥ $10^{24}$ dyne-cm). We did not use it in the analysis of each zone performed separately. It was used only in an aggregation of all 14 subduction zones (n=4609, M ≥ $10^{24}$ dyne-cm) into a single catalog. Our subduction zones differ from one to another by several geological/geophysical parameters (*Jarrard,*1986): slab dip, convergence rate, age of downgoing slab, length of the Benioff zone etc. Thus, some difference in the seismic regimes of these zones can be expected. However, we stress that they are all similar with respect to the subduction process dynamics.

To contrast with subduction zones, we consider as well the Mid Ocean Ridges (MOR). In this case, we have several seismic regimes governed by quite different dynamics that can be, in turn, split into two main classes: strike-slip events near transform zones, and normal faults (tension) near spreading zones. In order to retrieve events with these two source mechanisms, we used the diagram method elaborated in (*Kaverina et al.,*1996). As we shall see below, the differences in characteristics of plate dynamics are reflected in differences in the parameters of the distributions of event sizes corresponding to the different seismic regimes. We stress that this regionalization was performed before the statistical analysis and was fixed throughout the analysis, in order to avoid any possible bias.

## 3-The Generalized Pareto Distribution

As explained in (*Pisarenko and Sornette*, 2001), we model the seismic moment-frequency distribution by the so-called Generalized Pareto Distribution (GPD) defined as (*Embrechts et al.*, 1997)

(2)  $\quad G\,(y/\,\xi,\,s)\;=\;1 - (1 + \xi\,y\,/\,s)^{-1/\xi},$

where the two parameters $(\xi, s)$ are such that $-\infty < \xi < +\infty$ and $s > 0$. For $\xi \geq 0$, $y \geq 0$ and for $\xi < 0$, $0 \leq y \leq -s/\xi$. The GPD is a natural improvement of the power law distribution (1) and recovers it asymptotically for large y with the correspondence $\beta = 1/\xi$.

Let us first recall some facts about GPD (for more details see (*Embrechts et al.*, 1997; *Pisarenko and Sornette*, 2001). Let $\overline{F}\,(y)$ denote the tail of the DF $F(x)$ : $\overline{F}\,(y) = 1 - F(y)$. Other names for $\overline{F}\,(y)$ are the "complementary cumulative" distribution or "survivor" function. Let us denote by $n_u$ the number of those observations $y_1 \ldots y_n$ that exceed a threshold $u$ and by $x_1, \ldots, x_{nu}$ the observations decreased by $u : x_i = y_i - u$; $y_i > u$. The Gnedenko-Pickands-Balkema-de Haan theorem (*Embrechts et al.*, 1997) demonstrates the existence of a general approximation to the tail $\overline{F}\,(x)$ by a GPD as a tail estimator given by

$$\overline{F}\,(x + u) \cong \overline{G}\,(x\,/\hat{\xi}\,,\,\hat{s}\,) \times (n_u\,/\,n),$$

where

$$\overline{G}\,(x\,/\,\xi,\,s)\;=\;(1 + \xi\,x\,/\,s)^{-1/\xi}.$$

The estimates of the two parameters $\hat{\xi}$, $\hat{s}$ can be obtained through the Maximum Likelihood estimation (ML) (*Embrechts et al.*, 1997; *Pisarenko and Sornette*, 2001). The log-likelihood $l$ equals

$$(3) \qquad l = -n_u \ln s - (1 + 1/\xi) \sum_{1}^{n_u} \ln (1 + \xi x_i / s).$$

Maximization of the log-likelihood $l$ can be done numerically. The limit standard deviations of ML-estimates as $n \to \infty$ can be easily obtained (*Embrechts et al.*, 1997):

$$(4) \qquad \sigma_\xi = (1 + \xi) / \sqrt{n_u} \quad ; \qquad \sigma_s = s \sqrt{2(1 + \xi) / n_u} .$$

In practice, one usually replaces the unknown parameters in equations (3) by their estimates. It should be noted that the scale parameter $s = s(u)$ depends on the threshold u, while the shape parameter $\xi$ is in theory independent of u and solely determined by the DF F(x) of the data points. Thus, one can hope to find a reasonable GPD fit to the tail if it is possible to take a sufficiently high threshold u and to keep a sufficiently large number of excesses over it. Of course, this is not always possible.

The importance of the GPD lies in the fact that, according to the Gnedenko-Pickands-Balkema-de Haan theorem, the limit distribution of excesses over threshold u obeys the GPD (2), independently of the specific form of the DF of the original observations $y_1...y_n$. Our use of the GPD for the description of excesses stresses the tail of distributions. This fact is important for two reasons:

  - generally speaking, the distribution of excesses can be fitted more efficiently than the distribution over a large range;
  -the distribution of excesses puts the emphasis mainly on the seismic risk and the energy balance of earthquakes.

The shape parameter $\xi$ is of great interest in the analysis of the tails. When x becomes large and $\xi > 0$, the tail of the DF in equation ( 1 ) approaches a power function

$$\overline{G} (x / \xi, s) \cong (\xi x / s)^{-1/\xi} .$$

$1/\xi$ is therefore the exponent of the survivor distribution function. It corresponds asymptotically to the exponent $\beta$ for the Pareto law (1). Thus, the GPD is asymptotically scale invariant for $\xi > 0$. The parameterization of the tails of distribution by $\xi$ is more appropriate from a statistical point of view.

In the sequel, we apply this GPD-approach to the distribution of seismic moments M characterizing the energy release of earthquakes. In this case, the slope $b$ of the Gutenberg-Richter magnitude-frequency law is approximately proportional to the exponent $1/\xi$, with a coefficient of proportionality 2/3 : $b = 3/(2\xi)$ .

As an illustration of the application of the GPD approach to real catalogs, we fitted it to the aggregated sample of 14 subduction zones described above (n=4609, M$\geq 10^{24}$ dyne-cm), as well as to MOR events (n=926, M$\geq 10^{24}$, strike-slip; n=360, normal fault). The tail histograms of these samples are shown on Fig.1 together with the fitted GPD. Note first the considerable difference in the slopes for these 3 tails, which is reflected by very different values of the exponent and parameters of the GPD reported in the caption of figure1.

It is visually apparent from Fig.1 that the tail of the distribution of moments in the subduction zones contains about 20 extreme observations that deviate (visually) from the GPD curve. A "change point" occurs apparently somewhere near M = $5 \times 10^{27}$ dyne-cm (magnitude $m_w \cong 7.8$). A similar "change point" is seen in the tail of strike-slip MOR events, somewhere near M = $1.2 \times 10^{26}$ . In contrast, there is no visible "change point" in the tail of normal fault MOR events. Of course, a strict statistical test is needed to check the significance of these deviations and to characterize the "corner" magnitude. The development of such a test is the purpose of this paper and is now presented.

## 4-Test of the deviation from the GPD

Our method is based on the bootstrap approach (*Efron and Tibshirani*, 1986). The problem is divided into two parts. The first one consists in the statistical testing of the null hypothesis $H_0$ that the GPD is valid in the semi-infinite interval (u; ∞) for some u. The second part presented in section 5 includes the estimation of the "change point" if $H_0$ is rejected.

The statistical test of the hypothesis $H_0$ is constructed as described in details below. The steps of our approach are the following:

(i) We first rank order the seismic moments exceeding a lower threshold u: $y_l \geq ... \geq y_m$.

(ii) We perform a transformation from the extreme values $y_1 \geq ... \geq y_m$ into the variables $t_1 = \overline{F}(y_1) \leq ... \leq t_m = \overline{F}(y_m)$, where $\overline{F}(y)$ denotes the complementary cumulative GPD, i.e., the probability that a seismic moment is larger than y. This transformation converts variables y which vary extremely wildly into variables t with much more manageable fluctuations which are distributed approximately as m ordered random values with a uniform distribution in the interval (0,1). The existence of an approximation stems from the fact that we have to use the GPD with parameters ($\hat{\xi}$, $\hat{s}$) obtained from a statistical estimation procedure rather than use the absolutely exact DF. The mean value and the variance of $t_j$ for the exact DF are well known (Hajek and Sidak, 1967):

$$\mathbf{E}\ t_j = j / (N+1) ; \qquad \mathbf{Var}\ t_j = j\ (N-j+1) / (N+1)^2(N+2).$$

In order to construct a statistical test, we normalize the deviations $t_j$ :

$$\rho_j = (t_j - \mathbf{E}\ t_j) / (\mathbf{Var}\ t_j)^{1/2} .$$

These normalized deviations $\rho_j$ of the tail values $t_j$ for subduction zones and strike-slip MOR zones shown on Fig.1 are presented on Figs 2a and 2b. One can observe that 15 subduction events and 9 MOR events exceed one standard deviation. Comparing these graphs with the non-normalized ones shown in Fig 1, it is clear that the proposed normalization enhanced a lot the significance of the deviations.

One can also observe positive deviations for subduction events for small ranks (R≅100÷500) which exceed two standard deviations. This means that the GPD does not approximate the empirical distribution very well in this range. Perhaps, the lower threshold u should be increased.

Figs 2c and 2d present similar graphs for the case where the G-R distribution is used in the definition of $t_j$ and $\rho_j$. It is clear that this DF is less appropriate for normalization than the GPD: almost all normalized deviations exceed one standard deviation. Besides, a steady negative trend is present, which testifies to a poor approximation of the sample by the G-R law. Nevertheless, it is still clear that there is a change of behavior of the largest ranks, whose deviations are the strongest.

The results of the application of the normalization using the GPD for each separate zone taken individually are shown on Fig 3a-3l. We see that there is a suspicion for a bent down in graphs of Alaska, Mexico, New Hebrides, Solomon Isls and Taiwan. In section 5, we shall check their significance.

(iii) We now suggest a method for estimating the significance level of the observed deviations. In this

aim, we keep the $r$ first largest values of y and thus obtain variables $\rho_1, ..., \rho_r$ with approximately zero mean and unit variance. We take the sum of their squares $S_r = \rho_1^2 + ... + \rho_r^2$ as a measure of the deviation of the sample from the GPD.

(iv) We transform $S_r$ into a dimensionless statistic $\hat{\varepsilon}_r$:

$$(5) \qquad\qquad \hat{\varepsilon}_r = \Gamma(r/2; S_r/2),$$

where $\Gamma(a, x)$ is the incomplete Gamma function. This transformation makes it possible to compare the significance of the deviations for different values of $r$, and then to choose an optimal value of r for each catalog. If the normalized variables $\rho_i$ were standard independent Gaussian random values, then $\hat{\varepsilon}_r$ would give the probability of exceeding the value $S_r$ under the hypothesis $H_0$ ($\chi^2$-square distribution with r degrees of freedom). The smaller is $\hat{\varepsilon}_r$, the less probable is the hypothesis $H_0$ because a small $\hat{\varepsilon}_r$ means that the deviation of the sample from the GPD quantified by $S_r$ is so large that it cannot be accounted for by normal statistical fluctuations.

(v) For non-Gaussian variables with finite variance (as is the case here with the statistics of the variables $\overline{F}$), we estimate with any desired accuracy the statistical significance using the bootstrap method (see below).

(vi) We optimize the choice of $r$ by minimizing the value ($\hat{\varepsilon}_r$) over $r = 1...R$, where R is some a-priori chosen number (usually, we take R=20). Thus, the final decision statistic is

$$(6) \qquad\qquad \hat{\hat{\delta}}_{min} = \min_r(\hat{\varepsilon}_r) \quad .$$

The distribution of the statistic $\hat{\hat{\delta}}_{min}$ is estimated by the bootstrap method. We used in this estimating procedure 1000-10000 random trials with the parameters of the GPD fixed at their maximum likelihood estimates. Thus, we estimate the probability $\varepsilon$ of the random statistic $\hat{\hat{\delta}}_{min}$ under the hypothesis $H_0$ to be less than the observed sample value of $\min_r(\hat{\varepsilon}_r)$. As we noted already, the smaller the probability $\varepsilon$, the less probable is the hypothesis $H_0$. Note that, in our bootstrap procedure, we reproduce the whole algorithm of calculating the decision statistic $\hat{\hat{\delta}}_{min}$: we generate a random GPD sample with MLE estimates ($\hat{\xi}$, $\hat{s}$); then we estimate the GPD parameters by MLE and determine one value of $\hat{\hat{\delta}}_{min}$ in accordance with the method described above. Then, we repeat this procedure 1000 to 10000 times and estimate the probability that the observed value of $\hat{\hat{\delta}}_{min}$ would not be exceeded. This estimate characterizes the significance level of the hypothesis $H_0$.

The results of the application of our proposed technique to seismic zones are shown in Table 2. The decision statistic $\hat{\hat{\delta}}_{min}$ is significantly small (less than 5%) only for four regions: the aggregation of all 14 subduction zones, MOR stick-slip events, Solomon Isls, and Taiwan. For three zones, the deviations are on the borderline of significance: Alaska, Mexico, New Hebrides. A very distinct positive deviation is obtained for the sample including all 14 subduction zones. But even in the most favorable case of the subduction zones, the total number of clearly deviating extreme events is in the range 12-15, whereas in the other cases, this number is even smaller.

## 5-Determination of the "corner" magnitude

When the hypothesis $H_0$ is rejected, it is desirable to estimate the "corner seismic moment", or the "crossover point" $c_0$ defined as the value of seismic moment where the GPD becomes invalid, and a steeper decay starts to hold.

Any statistical estimation of the crossover point $c_0$ is necessarily very uncertain since it has to be based on a very small number of deviating events. This was justly noted already in [*Sornette et al.,*1996]. We describe below one of the most efficient statistical methods for the estimation of $c_0$ – the method of maximum likelihood. Unfortunately, even this powerful method cannot provide a reliable estimate of $c_0$ in most practical cases as it requires samples of sizes n=1000 and more, which are not available. We now present our method in order to quantify the amount of information that can be extracted from the data on the crossover point $c_0$.

We assume that the probability density f(x) is represented by two different dependencies on the intervals (u, $c_0$) and ($c_0$, ∞). In the first interval, we assume a GPD density whereas, on the second one, we assume some density φ(x) decreasing faster than the GPD. Thus, the PDF f(x) has the following form:

$$(7) \qquad f(x) = \begin{cases} a_1(1+\xi/s(x-u))^{-1-1/\xi} \; ; & u \le x \le c_0 \; ; \\ \\ a_2\varphi(x) \; ; & x \ge c_0 \end{cases}$$

The constants $a_1,a_2$ are chosen so that the density f(x) is continuous at the point $c_0$ and its integral over (u, ∞) equals unity. Modeling the second part of the tail is necessarily rather uncertain. In all practical situations, one has a very low number of observations supporting the estimation of the second branch and, as it was noted in [*Pisarenko and Sornette,* 2001], almost all possible models of the second branch are equally efficient, or, rather, equally inefficient. We shall try two variants of φ(x) with quite different behavior in the tail, power-like and exponential:

$$(8) \qquad \varphi_1(x) = \beta c_0^{\beta}/x^{1+\beta} \; ; \qquad \varphi_2(x) = 1/\alpha \, \exp(-(x-c_0)/\alpha) \; ; \quad x > c_0.$$

The parameter β (or α) is estimated together with the parameter $c_0$. We shall show that both these models result in essentially the same estimation accuracy of the crossover point $c_0$. In order to make the estimation problem more manageable, we assume that the parameters of the first branch are known (or can be estimated in a preliminary procedure with a good accuracy). Otherwise, we would have 4 unknown parameters whose estimation would be an almost insurmountable statistical problem given the scarcity of the data. Thus, only two parameters are assumed to be unknown: $c_0,\beta$ (or $c_0,\alpha$). They are estimated by the Likelihood Method. Note that both normalizing factors $a_1,a_2$ depend on the unknown parameters $c_0,\beta$ (or $c_0,\alpha$). It is necessary to note that the likelihood function with the PDF (7) is not differentiable, although it is continuous, so that its maximum can be easily found numerically. However, because of non-differentiability, it is impossible to use well-known formulae for limit variances/covariances of parameter estimates based on the Fisher's information matrix or the Hessian matrix. The only way for estimating these variances/covariances is provided by the bootstrap method (see below).

An illustration of an estimation of the crossover point $c_0$ and the corresponding model is shown on Fig 4 along with 10 realizations of random samples whose PDF satisfies eq.(7) with φ(x) = $\varphi_1(x)$. For comparison, we display the GPD branch of eq.(7) extended to infinity (without any crossover point). Note that the tail $\overline{F}$ (y) on Fig 4 is continuously differentiable since f(x) is continuous. A striking observation should be stressed: the deviation of the theoretical DF corresponding to eq.(7) and of the 10 random samples satisfying eq.(7) from the GPD without change of regime ($c_0$=infinity) starts much earlier than the true crossover point $c_0$. On fig 4, one can see a divergence between these curves starting approximately at $M=10^{27}$ whereas the true crossover is $c_0 = 1.4 \times 10^{28}$. The reason for this paradoxical result lies in the values of the coefficients $a_1$ and $a_2$, which are determined from the normalization of the global distribution (7). This condition of global normalization makes $a_1 < 1$ and thus explains the deviations of the model (7) from the pure GPD for values of the seismic moment smaller than the crossover $c_0$. This rather subtle fact should be kept in mind when the crossover is estimated "visually": such largely reported values for the crossover point log $c_0 \cong$ 27.37 ($m_w$ =7.5) for subduction zones and log $c_0 \cong 25.10$ ($m_w$ = 6.0) for MOR [*Pacheco et al.,* 1992; *Okal and Romanowicz,* 1994] (moments in dyne-cm) might be significantly underestimated (if one believes the visual estimations).

The likelihood function for the PDF (7) with the presence of the second branch (8) can be easily written, but we omit these trivial calculations. In accordance with the graphs on Figs 2a-d and 3a-l, one can expect the existence of a crossover point in the following zones: aggregation of all 14 subduction zones; strike-slip events in MOR; Alaska, Mexico, New Hebrides, Solomon Isls, Taiwan. For all the other zones, there is no sign of the existence of a crossover point, and the application of the maximum likelihood estimation is hopeless in such situations. One can only say that, if a crossover point exists, it should be much larger than the observed maximum of the corresponding sample. For the 7 zones mentioned above, we have applied our model (7) with the two variants (8) in the tail. The resulting Maximum Likelihood estimates of the corresponding parameters are shown in Table 3. We see that the MLE of $c_0$ obtained using the two models coincide for 5 zones and somewhat differ for only 2 zones (aggregation of 14 subduction zones, Solomon Isls). Such an agreement of the estimates derived from the two models confirms our opinion mentioned above that all models of the second branch of the PDF are almost equally efficient for the estimation of the crossover value $c_0$.

The estimates of the parameter $\beta$ (or $\alpha$) shown in Table 3 are extremely uncertain. Sometimes, the estimate of $\beta$ takes extremely large values (correspondingly, the estimates of $\alpha$ take very small values close to zero). Thus, the parameter $\beta$ (or $\alpha$) is such that the second branch of PDF in (7) practically shrunk down to zero just after $x = \max (X)$, where max(X) is the observed maximum of the sample. In these cases, our estimate of $c_0$ coincides with the MLE of the crossover value obtained using the truncated GPD [*Kijko and Sellevol,* 1989; *Kijko and Sellevol,* 1992; *Pisarenko et al.,*1996; *Kijko,*2001], whereas the estimate of $\beta$ (or $\alpha$) becomes senseless. Fig 5 shows a typical example of such an estimation. Here only 15 1/b-estimates from m=100 bootstrap samples have an intermediate value close to the true value 1/b=2/3, whereas 85 1/b-estimates are practically zero. Note that nevertheless lg($c_0$)-estimates do not exhibit such a "jump" in distribution for both sets of 1/b-estimates, although  lg($c_0$)-estimates for the former set are biased to the left with respect to the latter set. As we said above, any estimate of $c_0$ is rather uncertain. It is thus very important to characterize the statistical uncertainty of the estimate of $c_0$. For this purpose, we use again the bootstrap approach. We generate bootstrap samples $X^{(1)}...X^{(m)}$ of needed size n corresponding to the model (7),(8) with parameters fixed at their maximum likelihood estimate values $\hat{c}_0, \hat{\beta}$ (or $\hat{c}_0, \hat{\alpha}$). For the j-th bootstrap sample $X^{(j)}$, we determine $c_0^{(j)}, \beta^{(j)}$ (or $c_0^{(j)}, \alpha^{(j)}$) by MLE. Then, we estimate the bias and standard deviations of lg($c_0^{(j)}$), $\beta^{(j)}$ (or lg($c_0^{(j)}$), $\alpha^{(j)}$) from the "true" values $\hat{c}_0, \hat{\beta}$ (or $\hat{c}_0, \hat{\alpha}$). We generate bootstrap samples for the two PDF corresponding to $\varphi_1$ and $\varphi_2$ respectively. We then apply these two models to each of these bootstrap populations. We thus have 4 possible combinations: (power-like tail , power-like model), (power-like tail, exponential model), (exponential tail, power-like model), (exponential tail, exponential model). We have tried all these 4 cases. The corresponding biases and standard deviations are shown in Table 4 for a number of sample sizes and "true" parameter values. For these estimations, we used m=100 bootstrap samples for each variant. As it could be expected, the uncertainty of the estimation of $c_0$ (bias and standard deviation of lg($c_0^{(j)}$)) is practically the same for both models. For the sample size n=100, the bias and standard deviation are very large, in particular for larger lg($c_0$): the Mean-Square-Error = sqrt (bias$^2$ + std$^2$) is more than unity for all variants with $c_0 > 4 \times 10^{27}$. For n=250 and n=500, the Mean-Square-Error of lg($c_0$) is still high, and only for n > 1000 does the Mean-Square-Error become less than 0.4-0.5.

We have used this bootstrap method in order to estimate the bias and standard deviations for all 4 combinations of model/tail with parameters and sample sizes that were exactly equal to the MLE estimates obtained on the 7 samples tested for the possible existence of a crossover point (see section 4). The resulting Mean Square Errors are shown in Table 4. Comparing the two models (8) used for the estimation of $c_0$, we can conclude that the first one (the Pareto density) is preferable since it provides smaller MSE. Therefore, we used $c_0$-estimate provided by this model. As final estimates of the uncertainties in real samples we have taken the most conservative version (maximum value) of the uncertainty of the Pareto model for two versions of the "true" tail (Pareto and exponential). We were thus able to estimate the uncertainty of the MLE of $lg(c_0)$, see Table 3. Except for MOR events whose crossover point $c_0$ differs significantly from all subduction crossovers, we cannot affirm that crossovers differ significantly in various subduction zones, although their estimates vary within some limits. The relatively small sample sizes in separate zones do not allow us to obtain such a definite conclusion. However, it is not impossible that, say, in Solomon Isls and Taiwan, the crossovers can be less than in other subduction zones such as Alaska, Kurils, Mexico, South America, New Guinea, Tonga. At least, the MLE estimation of $c_0$ of the formers are less than the MLE estimations of $c_0$ of the latter. In order to obtain a more definite answer concerning distinct subduction zones, it would be necessary to double (or even to triple) the size of the existing catalogs. As to the global catalog of all subduction zones taken together, an estimate $lg(c_0) = 28.14 \pm 0.40$ ($m_W = 8.1 \pm 0.3$) can be accepted as reliable since it is based on a large sample of size $n=4609$.

If, for some zone, the hypothesis of validity of the GPD (or the G-R) on the semi-infinite interval $(u; \infty)$ is rejected, and the MLE-estimate of the crossover point $c_0$ has been derived, then it is natural to re-estimate the GPD form parameter $\xi$ (or the G-R slope parameter b) using only the data from the interval $(u; c_0)$ rather than from the interval $(u; \infty)$. On the interval $(u; c_0)$, the likelihood function is used for the corresponding distribution truncated from both sides: this results in a new normalizing constant depending on the unknown parameters. We have carried out such a re-estimation for the 7 zones whose deviation from the unlimited GPD was found significant (see Table 3). The results of this re-estimation are compared in Table 5 with the estimates obtained on the semi-infinite interval $(u; \infty)$. We observe that the corresponding differences of the estimates are not always negligible. Sometimes, they reach 10%. Thus, in the case when a finite crossover point $c_0$ has been derived with a reasonable reliability, it is safer to estimate the form parameters of the GPD (or the G-R) using DF truncated from both sides.

# 6-DISCUSSION AND CONCLUSIONS

We have started our analysis from the observation that the Generalized Pareto Distribution (2) (GPD) provides a satisfactory approximation of the tails of the distributions of the seismic energy released by earthquakes [*Pisarenko and Sornette*, 2001]. The justification for the use of the GPD is that it offers an improvement over the simple Pareto power law (1) as it is rigorously based on the Gnedenko-Pickands-Balkema-de Haan theorem, that shows that the GPD is the universal distribution of sizes conditioned to exceed a threshold, in the limit where this threshold becomes large, independently of the specific distribution of the unconditional values. The GPD has also a power law tail and thus recovers exactly a power law, asymptotically.

Even if the GPD works well in the intermediate range of seismic catalogs, there is always the possibility that, at the extreme end of the range of sizes, some deviation from the GPD may occur. Since, as we said, the use of the GPD is warranted asymptotically by the Gnedenko-Pickands-Balkema-de Haan theorem, such a deviation would signal a possible change of physics and the existence of new mechanisms that could control the GDP parameters.

The very important question of the possible existence of crossover point $c_0$ in the magnitude frequency law has thus been studied in this paper. First of all, a quantitative statistical test for the presence of a crossover $c_0$ has been introduced, based on the transformation of the ordered sample of seismic moments into a series with uniform distribution under condition of no crossover. The subsequent use of the bootstrap method has allowed us to estimate the statistical significance of the null hypothesis $H_0$ (absence of crossover). If $H_0$ is rejected, we have shown how to address the next question, which is to estimate the crossover $c_0$. We found that, for a reliable estimation of $lg(c_0)$, a rather high minimum sample size is necessary that can be evaluated approximately as $n \cong 1000$: such sample size would provide Mean-Square-Errors (MSE) of the estimate of $lg(c_0)$ no more than 0.4-0.5; for sample size $n=500$ and less, the MSE can reach 0.5 to 1.1. Therefore, the estimation of the crossover $c_0$ is possible only for very large geographical areas with numerous events.

For the catalog obtained by aggregating 14 subduction zones of the Circum Pacific Seismic Belt, our estimate of the crossover point is $\lg(c_0) = 28.14 \pm 0.40$ ($c_0$ in dyne-cm), corresponding to a crossover magnitude $m_W = 8.1 \pm 0.3$. For separate subduction zones, the corresponding estimates are much more uncertain (see Table 5), so that the null hypothesis of an identical crossover for all subduction zones cannot be rejected. However, it is possible that this conclusion is due only to the insufficient sample sizes in the separate zones. Our conclusion does not exclude a spatial variation of the crossover value $c_0$. For the 14 subduction zones, the 4 largest earthquakes turned out to be beyond the ML-estimate $\lg(c_0) = 28.14$. Thus, they can be considered as deviating significantly from the GPD tail:

$M = 3.00 \times 10^{28}$, 04.10.1994, $\lambda = 43.71$; $\varphi = 147.33$; Kurils;

$M = 2.41 \times 10^{28}$, 17.02.1996, $\lambda = -.95$; $\varphi = 17.03$; New Guinea;

$M = 1.69 \times 10^{28}$, 12.12.1979, $\lambda = 1.62$; $\varphi = -79.34$; South America;

$M = 1.39 \times 10^{28}$, 22.06.1977, $\lambda = -22.91$; $\varphi = -175.74$; Tonga.

Here $\lambda$ and $\varphi$ are the latitude and the longitude correspondingly. The evidence demonstrated here of deviations from the Pareto or GPD is usually related to the finite thickness of the seismogenic layers (although no direct evidence of this statement is demonstrated in the existing literature (Sornette et al., 1996; Main, 2000)). With our new statistical approach, we find that the crossover magnitude is $m_W = 8.1 \pm 0.3$ for subduction zones. Such a large value makes it difficult to associate it directly with a seismogenic thickness as proposed by many different authors in the past. It may point to the concept that the non-radiating part of the lower crust may participate significantly in the mechanical localization and stress relaxation processes associated with an earthquake, according to their visco-elasto-plastic rheological behavior. In other words, the localization of strong shear could propagate significantly in the lower crust and upper mantle, thus increasing the effective size beyond which one should expect a change of regime. While this idea is not new, our statistical tests leading to such a large value of the crossover magnitude may be one of its clearest signatures.

The strong statistical significance of the deviations that we have demonstrated above $c_0$ justifies the quest for a parametric representation of the second branch of the PDF describing these deviations (Kagan and Schoenberg, 2001). However, the number (of the order of 15) of events in the new regime is not sufficient for establishing any functional form of this PDF. As we already mentioned, there has been many attempts to fit the tail of the magnitude-frequency law by several parametric families. However, the "visual" as well as statistical quality of the fits with these families are almost the same. The situation is even more uncertain for separate regional catalogs because of the smaller number of observations.

What statistical recommendations can be suggested concerning the seismic hazard (seismic risk) assessment and on related problems? Of course, when the sample size is small, no statistical method can help in a definitive way, but still some cautionary measures can be recommended. First of all, it is desirable to use several competing models of the second branch of the tail and to compare them. By inserting the extreme parameter values of the confidence domain into a fitted tail function, one can compare the resulting difference of probabilities. The bootstrap approach can be very useful in this situation. Statistical estimation or hypothesis testing are easily modeled by the bootstrap method even for small samples. Sometimes, generating artificial samples and simple visual inspection can help in drawing conclusions.

**ACKNOWLEDGEMENTS**
The authors are thankful to P. Bird, H. Houston and Y.Y. Kagan for useful discussions. This work was supported partially by INTAS grant 99-00099, RFRF grant 99-05-64924, and by ISTC Project 99-1293 (VP) and by NSF-DMR99-71475 and the James S. Mc Donnell Foundation 21st century scientist award/studying complex system (VP and DS).

# REFERENCES

Bird, P., Kagan, Y.Y. and D.D. Jackson (2000) Plate tectonics and earthquake potential of spreading ridges and oceanic transform faults, UCLA preprint.

Cornell, C.A. (1994) Statistical analysis of maximum magnitudes, in: The Earthquakes of Stable Continental Regions, vol.1, edited by J. Schneider, pp.5/1-5/27, Electr. Power Res. Inst., Palo Alto, Calif.

Dargahi-Noubary, G.R. (1986) A method for predicting future large earthquakes using extreme order statistics, Phys. of the Earth and Planetary Interior, 42, 241-245.

Dargahi-Noubary, G.R. (2000) Statistical Methods for Earthquake Hazard Assessment and Risk Analysis, Nova Science Publishers, Inc., N.Y.

DeMets, C., R.G. Gordon, D.F. Argus, and S. Stein (1990) Current plate motions, Geophys. J. Int. 101, 425-478.

DeMets, C., R.G. Gordon, D.F. Argus, and S .Stein (1990) Effect of recent revisions to the geomagnetic reversal time scale on estimate of current plate motions, Geophys. Res. Lett., 21, № 20, 2191-2194.

Dziewonski, A.M., G. Ekstrom, and M.P. Salganik (1994) Centroid-moment tensor solutions for January-March, 1994, Phys. Earth Planet. Inter., 86, 253-261.

Efron, B., and R.Tibshirani (1986) Bootstrap method for standard errors, confidence intervals and other measures of statistical accuracy, Statistical Science, 1, 54-77.

Embrechts, P., C.P. Kluppelberg, and T. Mikosh (1997) Modelling Extremal Events, Springer-Verlag, Berlin, 645 pp.

Flinn, E.A., E.R. Engdahl, and A.R. Hill (1974) Seismic and Geographical Regionalization, Bull. Seismol. Soc. Am., 64, 771-792.

Gutenberg, B., and C.F. Richter (1954) Seismicity of the Earth and Associated Phenomena, 310 pp.

Hajek, J., and Z. Sidak (1967) Theory of Rank Tests, Academia Publ. House, Prague, 375 pp.

Jarrard, R.D. (1986) Relations Among Subduction Parameters, Rev. of Geophys., 24, 217-284.

Kagan, Y.Y. (1994) Observational Evidence for Earthquakes as a Nonlinear Dynamic Process, Physica D, 77, 160-192.

Kagan, Y.Y. (1997) Seismic Moment-frequency Relation for Shallow Earthquakes: Regional Comparison, J. Geophys. Res., 102, 2835-2852.

Kagan, Y.Y. (1999) Universality of Seismic Moment-frequency Relation, Pure appl. Geophys., 155, 537-573.

Kagan, Y.Y., and F. Schoenberg (2001) Estimation of the upper cutoff parameter for the tapered Pareto distribution, J.Appl.Probab., 38 A, 901-918.

Kanamori, H. and D. L. Anderson (1975) Theoretical basis of some empirical relations in seismology,, Bull. Seismol. Soc. Am. 65, 1073-1095.

Kaverina, A.N., A.V. Lander, and A.G. Prozorov (1996) Global creepex distribution and its relation to earthquake source geometry and tectonic origin, Geophys. J. Intern., 125, 249-265.

Kijko, A. (2001) Statistical Estimation of Maximum Regional Earthquake Magnitude $m_{max}$ ,12-th European Conference on Earthquake Engineering, Paper Reference FW:022, Elsevier Science Ltd.

Kijko, A., and M.A. Sellevol (1989) Estimation of earthquake hazard parameters from incomplete data files. Part I, Utilization of extreme and complete catalogues with different threshold magnitudes, Bull.Seism.Soc.Am, 79, 645-654.

Kijko, A., and M.A. Sellevol (1992) Estimation of earthquake hazard parameters from incomplete data files. Part II, Incorporation of magnitude heterogeneity, Bull.Seism.Soc.Am, 82, 120-134.

Knopoff, L., and Y. Y. Kagan (1977) Analysis of the theory of extremes as applied to earthquake problems, J. Geophys. Res., 82, 5647-5657.

Kronrod, T.L. (1984) Seismicity parameters for the main high-seismicity regions of the world, Vychislitel'naya Seismologiya, Nauka, Moscow, 17, 36-58 (in Russian); (Computational Seismology, Engl. Transl., 17, 35- 54, 1984).

Laherrere, J., and D. Sornette (1998) Stretched exponential distributions in Nature and economy: "fat tails" with characteristic scales, Eur. Phys. J. B 2, 525-539.

Main, I. (1992) Earthquake Scaling, Nature, 357, 27-28.

Main, I., (1996) Statistical Physics, Seismogenesis, and Seismic Hazard, Reviews of Geophysics, 34, 433-462.

Main, I. (2000) Apparent breaks in scaling in the earthquake cumulative frequency-magnitude distribution: Fact or artifact? Bull. Seismol. Soc. Am., 90, 86-97)

Main, I., and P.W. Burton (1984) Information Theory and the Earthquakes Frequency-magnitude Distribution, Bull. Seismol. Soc. Am., 74, 1409-1426.

Molchan, G.M., T.L. Kronrod, O.E. Dmitrieva, and A.K. Nekrasova (1996) Multiscale Model of Seismicity Applied to Problems of Seismic Risk: Italy, Vychislitel'naya Seismologiya, Nauka, Moscow, 28, 193-224 (in Russian).

Molchan, G.M., T.L. Kronrod, and G.F. Panza (1997) Multiscale Seismicity Model for Seismic Risk, Bull. Seismol. Soc. Am., 87, 1220-1229.

Ogata,Y., M. Imoto, and K. Katsura (1991) 3-D spatial variation of b-values of magnitude-frequency distribution beneath the Kanto District, Japan, Geophys. J. Intern., 104, 135-146.

Okal, E.A., and B.A. Romanowicz (1994) On the variation of b-values with earthquake size, Phys. Earth Planet. Inter., 87, 55-76.

Pacheco, J.F., and L. Sykes (1992) Seismic moment catalog of large, shallow earthquakes, 1900-1989, Bull. Seismol. Soc. Am., 82, 1306-1349.

Pacheco, J.R., C.H. Scholz, and L.R. Sykes (1992) Changes in frequency-size relationship from small to large earthquakes, Nature, 335, 71-73.

Pisarenko, V.F., A.A. Lyubushin, V.B. Lysenko, T.V. Golubeva (1996) Statistical Estimation of Seismic Hazard Parameters: Maximum Possible Magnitude and Related Parameters, Bull. Seism. Soc. Amer., 86, 691-700.

Pisarenko, V.F. and Sornette, D. (2001) Characterization of the frequency of extreme events by the generalized Pareto distribution, PAGEOPH submitted. (preprint at http://arXiv.org/abs/cond-mat/0011168)

Romanowicz, B., (1994) A reappraisal of large earthquake scaling – Comment, Bull. Seismol. Soc. Am., 84, 1765-1676.

Romanowicz, B., and J.B. Rundle (1993) On scaling relations for large earthquakes, Bull. Seismol. Soc. Am., 83, 1294-1297.

Rundle, J. B. (1989) Derivation of the complete Gutenberg-Richter magnitude-frequency relation using the principle of scale invariance, J. Geophys. Res. 94, 12,337-12,342.

Sornette, D., and A. Sornette (1999) General theory of the modified Gutenberg-Richter law for large seismic moments, Bull. Seism. Soc. Am., 89, 1121-1130.

Sornette, D., L. Knopoff, Y.Y. Kagan, and C.Vanneste (1996) Rank-ordering statistics of extreme events: application to the distribution of large earthquakes, J. Geophys. Res., 101, 13883-13893.

Young, J.B., B.W.Presgrave, H.Aichele, D.A.Wiens, and E.A.Flinn (1996) The Flinn-Engdahl regionalization scheme: the 1995 revision, Phys. Earth Planet. Inter., 96, 223-297.

**Figure captions:**

Figure 1: Empirical tail histogram and the corresponding fitted Generalized Pareto Distribution (GPD) for (1) the aggregated sample of 14 subduction zones (n=4609, M≥$10^{24}$ dyne-cm), (2) the MOR events (n=926, M≥$10^{24}$, strike-slip) and (3) the MOR events (n=360, normal fault). The parameters of the GPD fit are respectively:

$\xi = 1.517 \pm 0.033$ ; s = $20.95 \pm 0.57$ (1); $\xi = .937 \pm 0.056$ ; s = $29.82 \pm 1.64$ (2); $\xi = .659 \pm 0.075$ ; s = $10.64 \pm 0.41$ (3).

Figure 2: Normalized deviations $\rho_j$ of the tail values for (a) 14 subduction zones, (b) strike-slip MOR events. The right panels are magnifications of the left panels. The parameters of the GPD fit are the same as on Figure 1. (c-d) Same as (a-b) for the case where the GR distribution is used in the definition of $t_j$ and $\rho_j$. The parameters of the GR fit are b = $.582 \pm .009$ (14 subduction zones); b = $.617 \pm .021$ (MOR, strike-slip).

Figure 3: Normalized deviations $\rho_j$ of the tail values for each separate zone taken individually. GPD was used in the normalization procedure described in the text using the MLE estimates of ($\xi$, s)-parameters. (a) Alaska; (b) Japan; (c) Kamchatka; (d) Kurils; (e) Marianas; (f) Mexico; (g) New Hebrides; (h) Solomon; (i) South America; (j) Sunda; (k) Taiwan; (l) Tonga.

Figure 4: Illustration of the crossover point $c_0$ in model (7) with $\varphi(x) = \varphi_1(x)$ (power-like second branch, noted 2), along with 10 realizations of random samples whose PDF satisfies eq.(7). For comparison, the GPD branch of eq.(7) extended to infinity (without any crossover point) is also shown as the thick straight line noted 1.

Figure 5: ML-estimates of parameters ( lg $c_0$, 1/b) in model (7) with $\varphi(x) = \varphi_1(x)$ (power-like second branch) for 100 bootstrap samples with true parameters: lg $c_0$ = 27.7; 1/b = .667. GPD parameters of the first branch in eq.(7) were fixed at $\xi = 1.5$; s = 20; sample size n = 200.

*Table 1. Parameters of catalogs used in the analysis; taken out of the*
*Harvard CMT catalog, 1977-2000.*

| Region | Reference Position Lat., Long. | Number of events M$\geq 10^{24}$ dyne-cm | max M / $10^{27}$ |
|---|---|---|---|
| Alaska | 60; -152 | 332 | 10.4 |
| Japan | 36; 140 | 199 | 4.9 |
| Kamchatka | 53; 162 | 173 | 5.3 |
| Kuril Isls | 45; 152 | 257 | 30.0 |
| Mariana Isls | 17; 148 | 261 | 5.2 |
| Mexico | 16; -100 | 276 | 11.5 |
| New Guinea | -6; 150 | 128 | 24.1 |
| New Hebrides | -17; 167 | 439 | 4.8 |
| Solomon Isls | -7; 155 | 474 | 4.6 |
| South America | -20; -70 | 363 | 16.9 |
| South Sandwich Isls | -58; -24 | 130 | 0.6 |
| Sunda | -2; 98 | 422 | 7.3 |
| Taiwan | 10; 125 | 524 | 4.1 |
| Tonga | -22; -174 | 631 | 13.9 |
| Aggregation of all 14 subduction zones | | 4609 | 30.0 |
| **Midocean ridges** | | | |
| Spreading segments (normal fault, tension) | | 360 | .174 |
| Transform segments (strike-slip) | | 926 | .556 |

*Table 2. Significance levels of the hypothesis $H_0$ : unbounded GPD.*

| Region | $\hat{\delta}_{min}$ | $P\{\delta_{min} \leq \hat{\delta}_{min}\}$ |
|---|---|---|
| Aggregation of 14 subduction zones | $\hat{\delta}_{min} = \varepsilon_7 = 10^{-15}$ | 0 |
| MOR strike-slip | $\hat{\delta}_{min} = \varepsilon_2 = .0020$ | 2.3% |
| Alaska | $\hat{\delta}_{min} = \varepsilon_6 = .0793$ | 7.7% |
| Mexico | $\hat{\delta}_{min} = \varepsilon_1 = .0877$ | 7.2% |
| New Hebrides | $\hat{\delta}_{min} = \varepsilon_1 = .0572$ | 7.9% |
| Solomon Isls | $\hat{\delta}_{min} = \varepsilon_1 = 4.6 \times 10^{-4}$ | 1.2% |
| Taiwan | $\hat{\delta}_{min} = \varepsilon_6 = 1.9 \times 10^{-5}$ | 0.67% |

*Table 3. The MLE estimates of log $c_0$ provided by two models (the cross-over moment $c_0$ is expressed in dyne-cm)*

| Region | Pareto model | | Exponential model | |
|---|---|---|---|---|
| | MLE of lg $c_0$ | MLE of $\beta$ | MLE of lg $c_0$ | MLE of $\alpha$ |
| Aggregation of 14 subduction zones | 28.14 ±.40 | 2.27 | 28.38 ± .63 | $5.2 \times 10^4$ |
| MOR strike-slip | 26.47 ± .20 | 2.73 | 26.61 ± .37 | 931 |
| Alaska | 28.02 ± .47 | $6.3 \times 10^8$ | 28.02 ± .50 | $1.3 \times 10^{-8}$ |
| Mexico | 28.06 ± .43 | $8.3 \times 10^{13}$ | 28.06 ± .42 | $9.3 \times 10^{-9}$ |
| New Hebrides | 27.68 ±.35 | $6.2 \times 10^{12}$ | 27.68 ± .34 | $4.0 \times 10^{-8}$ |
| Solomon Isls | 27.66 ± .36 | $1.3 \times 10^4$ | 27.40 ± .31 | $1.2 \times 10^4$ |
| Taiwan | 27.61 ± .42 | $6.2 \times 10^{12}$ | 27.61 ± .41 | $3.0 \times 10^{-9}$ |

*Table 4. Mean Square Errors of MLE estimates of the log-crossover point lg $c_0$ by two models (artificial examples).*

| Sample | Pareto model | | | | | Exponential model | | | | |
|---|---|---|---|---|---|---|---|---|---|---|
| | n=100 | n=250 | n=500 | n=1000 | n=4600 | n=100 | n=250 | n=500 | n=1000 | n=4600 |
| lg $c_0$ = 27.60 | | | | | | | | | | |
| Pareto tail | 0.80 | 0.51 | 0.47 | 0.44 | 0.31 | 0.71 | 0.43 | 0.38 | 0.52 | 0.74 |
| Exp tail | 0.66 | 0.32 | 0.25 | 0.26 | 0.19 | 0.67 | 0.42 | 0.37 | 0.32 | 0.27 |
| lg $c_0$ = 28.15 | | | | | | | | | | |
| Pareto tail | 1.07 | 0.74 | 0.54 | 0.50 | 0.40 | 1.02 | 0.64 | 0.35 | 0.26 | 0.63 |
| Exp tail | 1.01 | 0.66 | 0.43 | 0.23 | 0.10 | 1.16 | 0.74 | 0.62 | 0.44 | 0.19 |
| lg $c_0$ = 28.48 | | | | | | | | | | |
| Pareto tail | 1.44 | 1.00 | 0.68 | 0.54 | 0.37 | 1.45 | 0.82 | 0.47 | 0.36 | 0.50 |
| Exp tail | 1.35 | 0.82 | 0.58 | 0.35 | 0.11 | 1.44 | 1.04 | 0.73 | 0.53 | 0.14 |

*Table 5. Comparison of ML-estimates obtained on the interval (u₀ ; ∞) with ML-estimate obtained on the interval (u₀ ; c₀) for the GPD and the GR form parameters; c₀ –values are taken from Table3; lower threshold u = 10²⁴ dyne-cm.*

| Region | GPD | | Gutenberg-Richter | |
|---|---|---|---|---|
| | MLE of $\xi$ on $(u ; \infty)$ | MLE of $\xi$ on $(u ; c_0)$ | MLE of $b$ on $(u ; \infty)$ | MLE of $b$ on $(u ; c_0)$ |
| 14 subduction zones | $1.517 \pm .033$ | $1.559 \pm .037$ | $.582 \pm .009$ | $.570 \pm .009$ |
| MOR strike-slip | $.933 \pm .055$ | $1.045 \pm .077$ | $.622 \pm .021$ | $.531 \pm .025$ |
| ALASKA | $1.557 \pm .129$ | $1.637 \pm .148$ | $.581 \pm .032$ | $.564 \pm .034$ |
| MEXICO | $1.894 \pm .159$ | $2.084 \pm .208$ | $.493 \pm .030$ | $.465 \pm .032$ |
| NEW HEBRIDES | $1.488 \pm .108$ | $1.571 \pm .131$ | $.522 \pm .025$ | $.492 \pm .028$ |
| SOLOMON | $1.480 \pm .104$ | $1.635 \pm .140$ | $.511 \pm .024$ | $.464 \pm .027$ |
| TAIWAN | $1.548 \pm .102$ | $1.690 \pm .129$ | $.568 \pm .025$ | $.539 \pm .027$ |

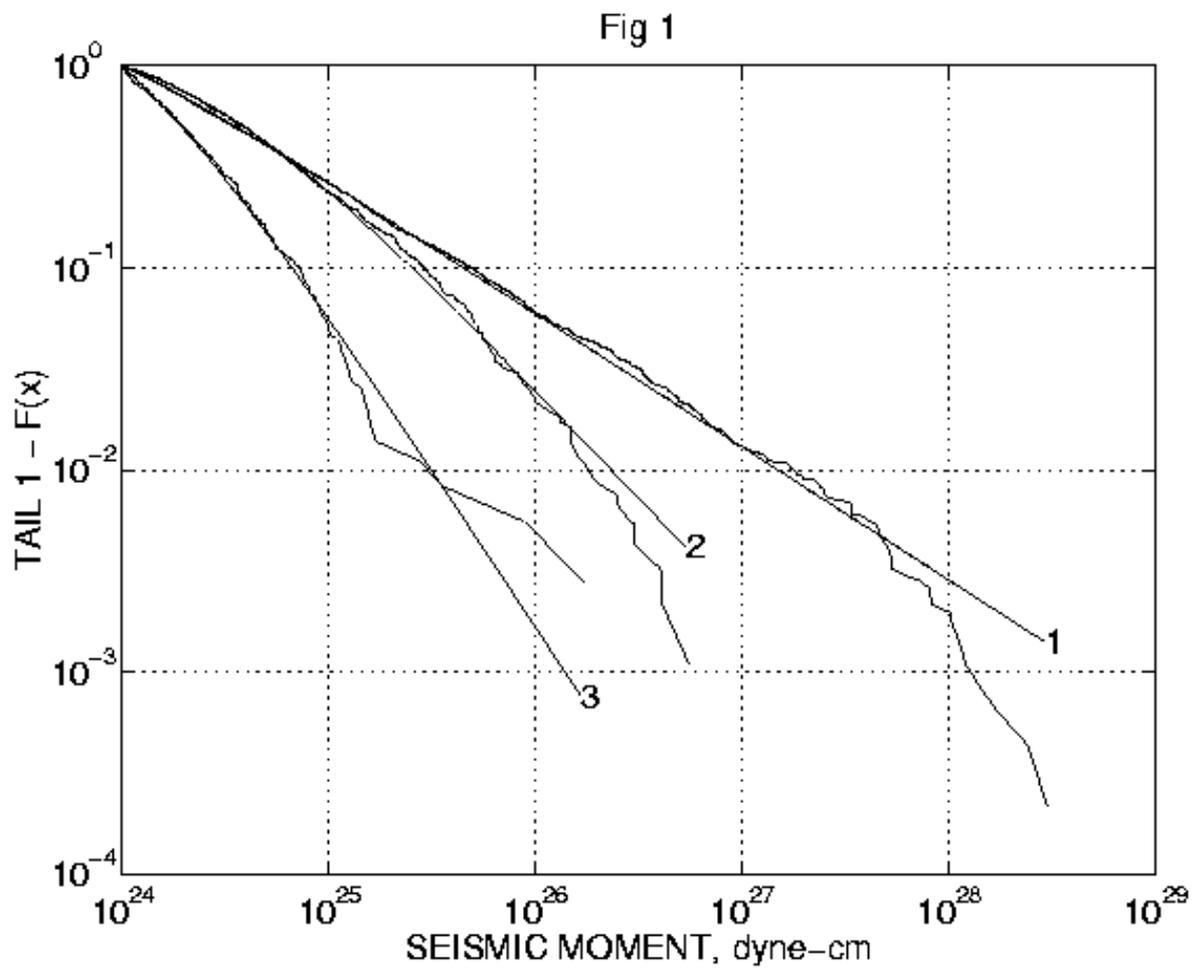

Fig 1

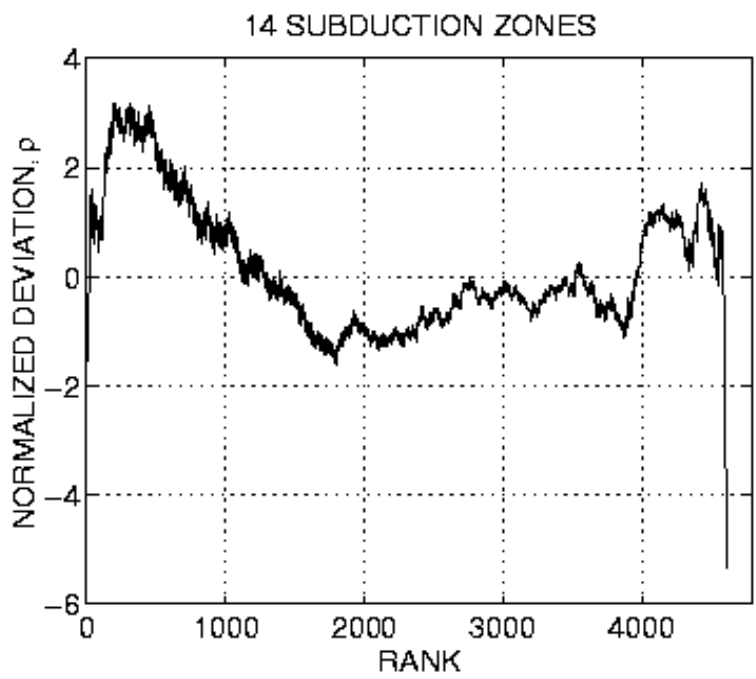

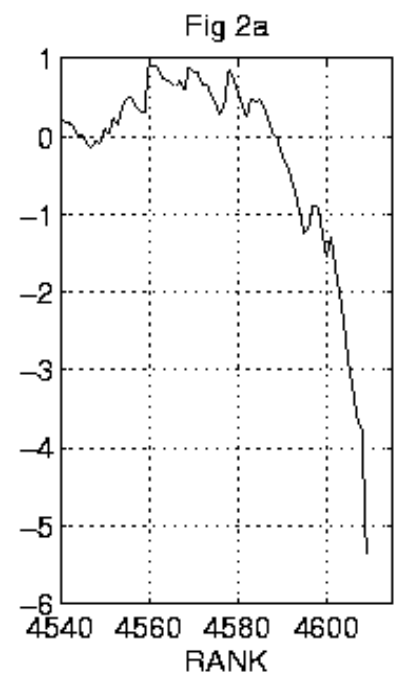

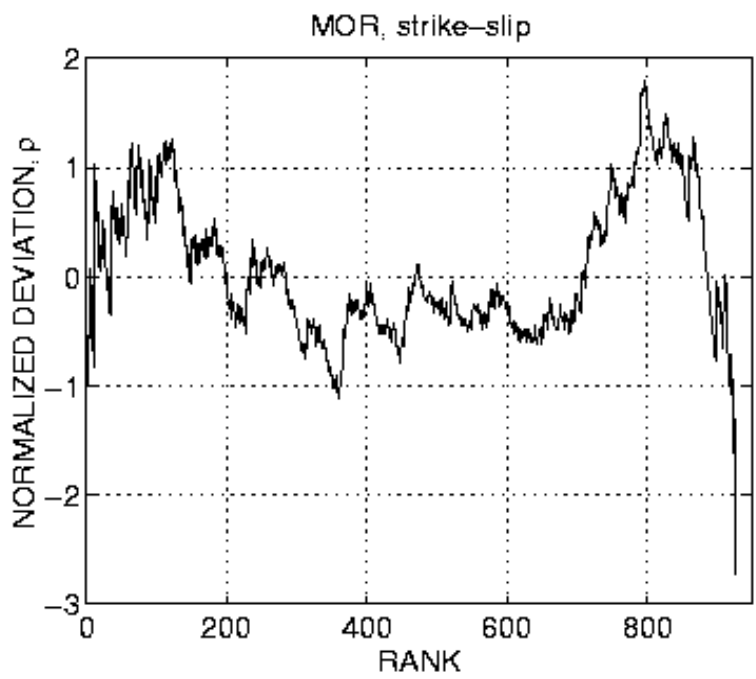
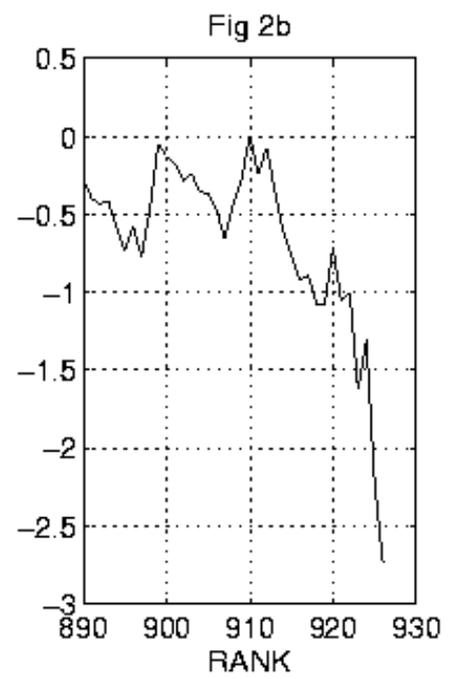

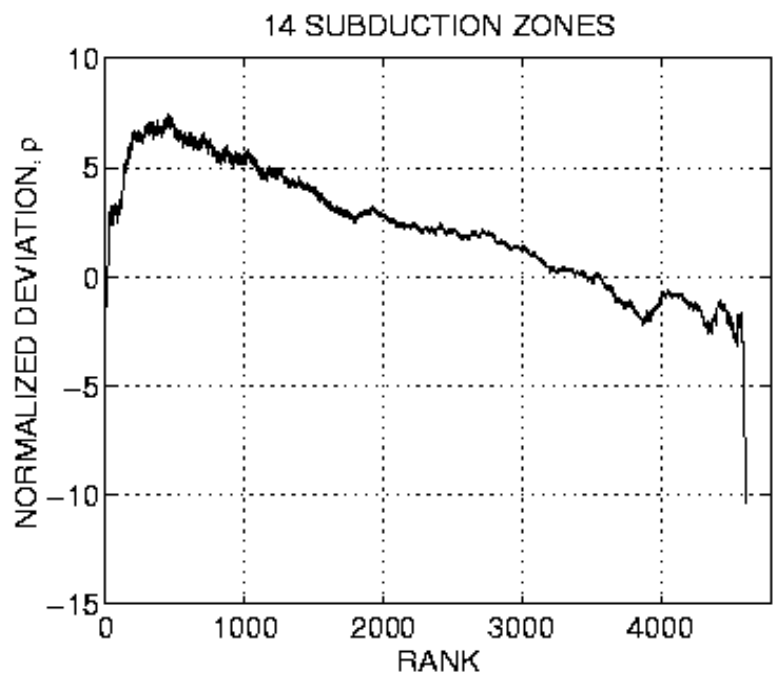

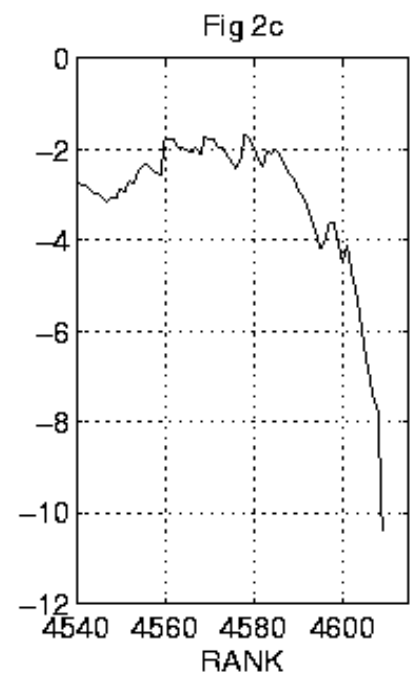

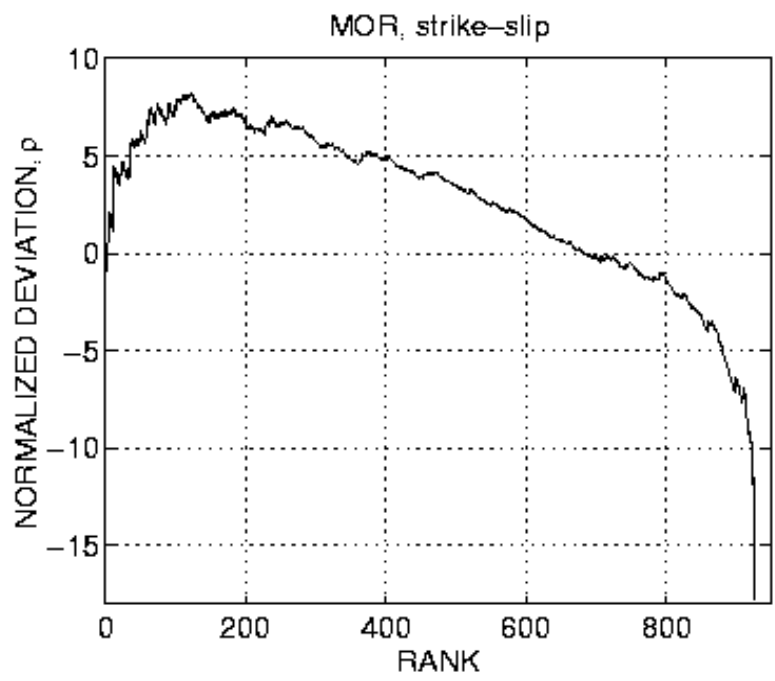
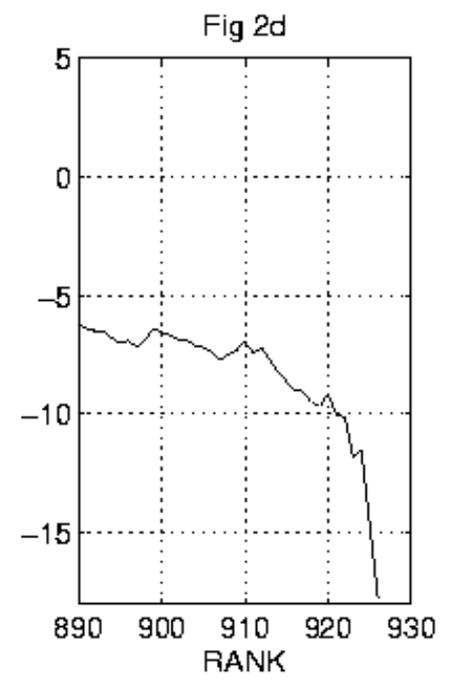

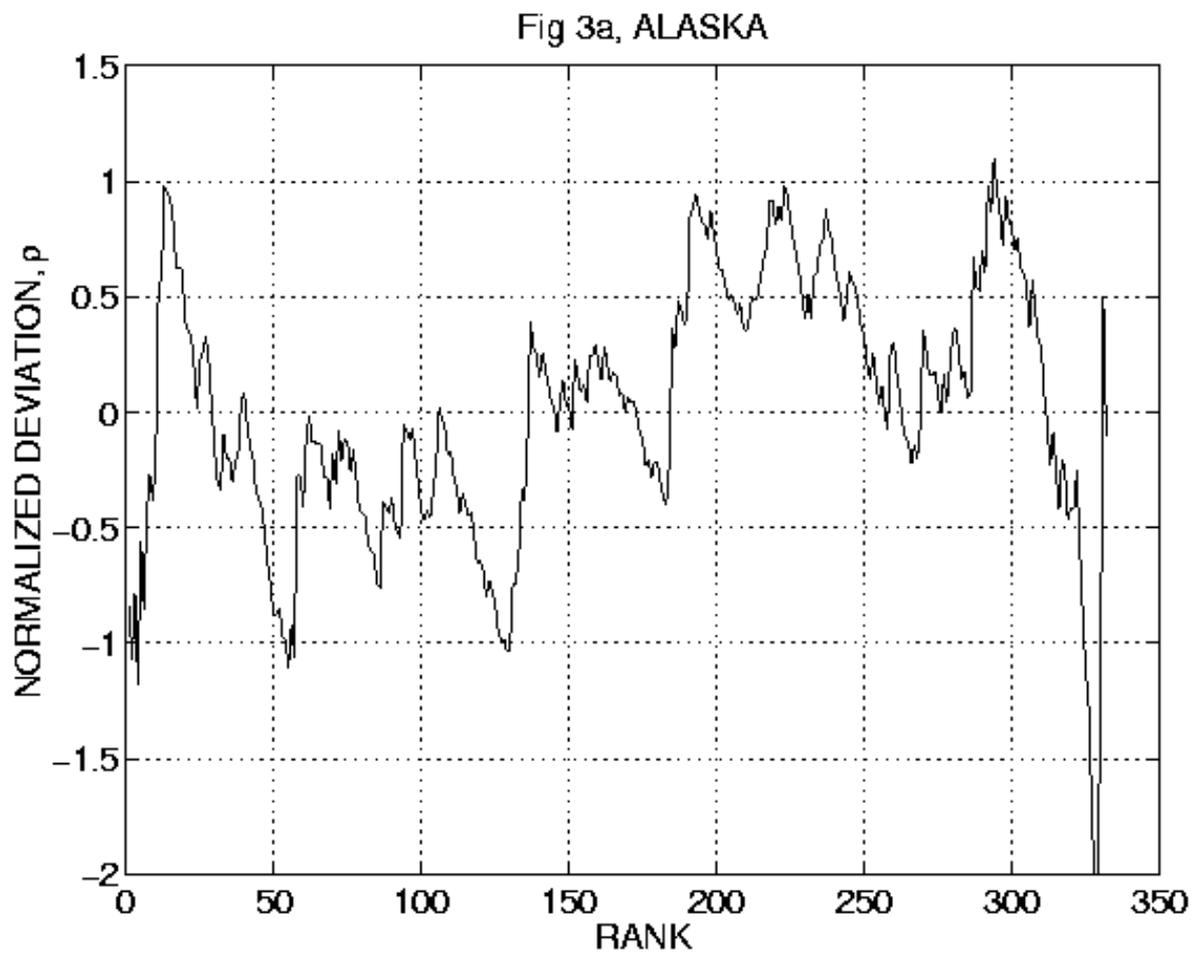

Fig 3a, ALASKA

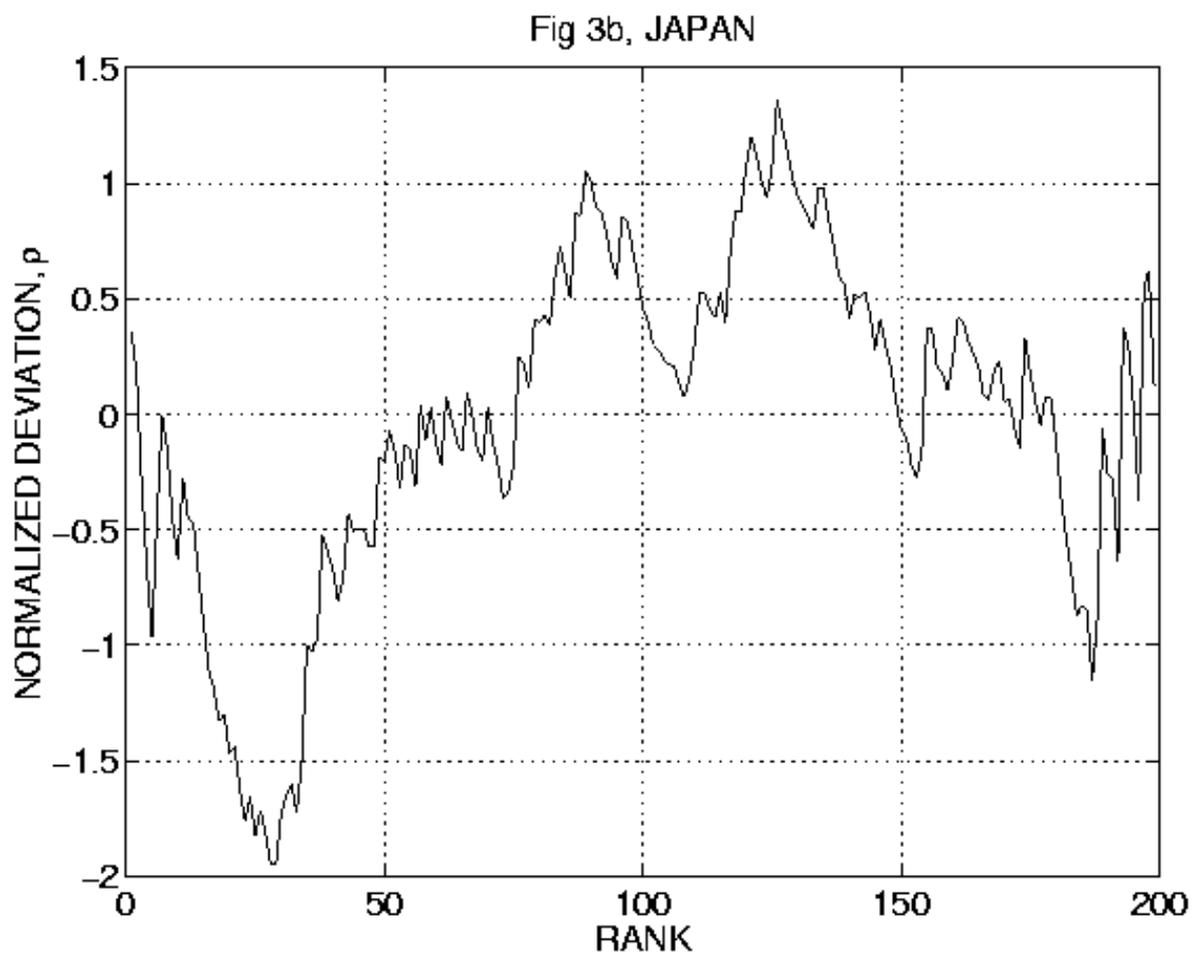

Fig 3b, JAPAN

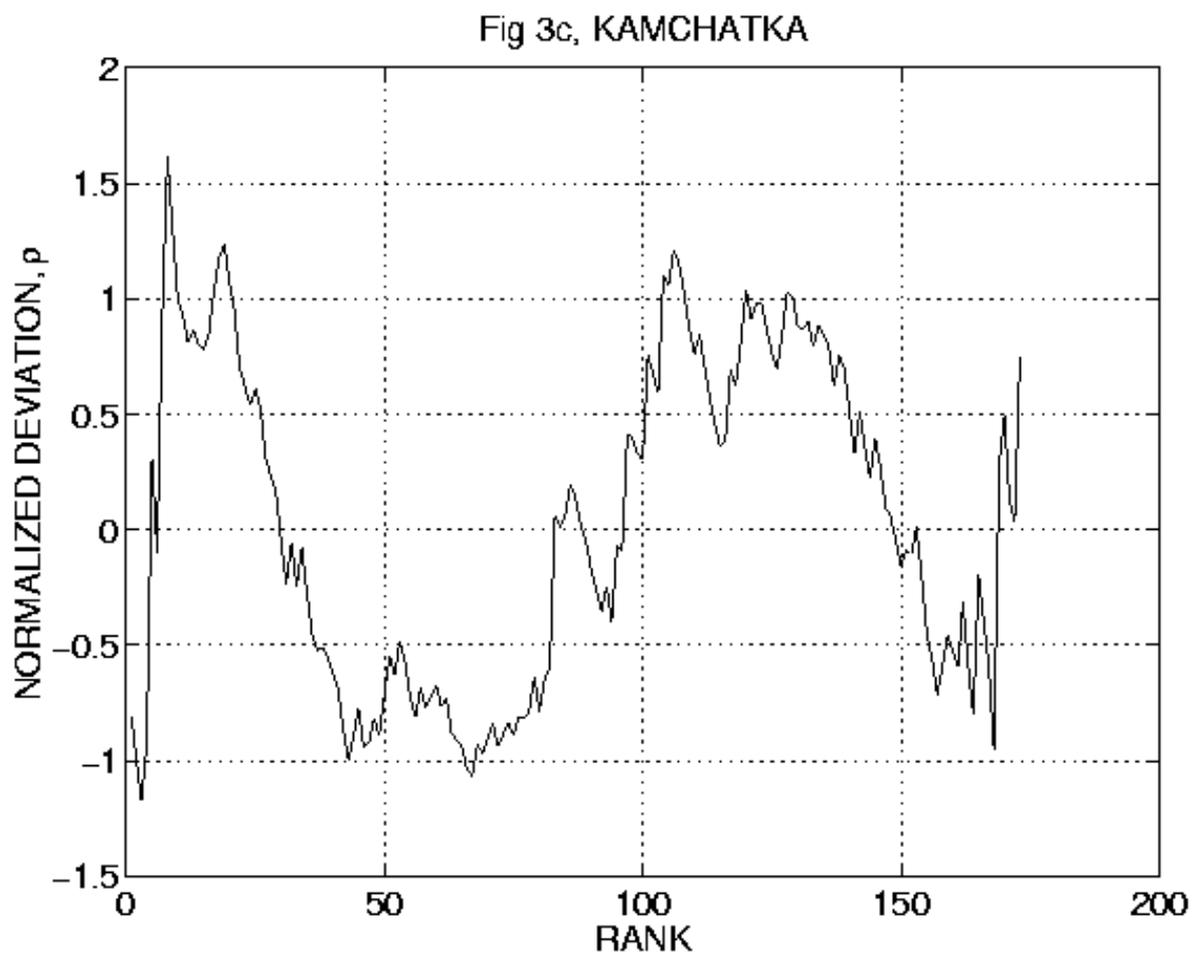

Fig 3c, KAMCHATKA

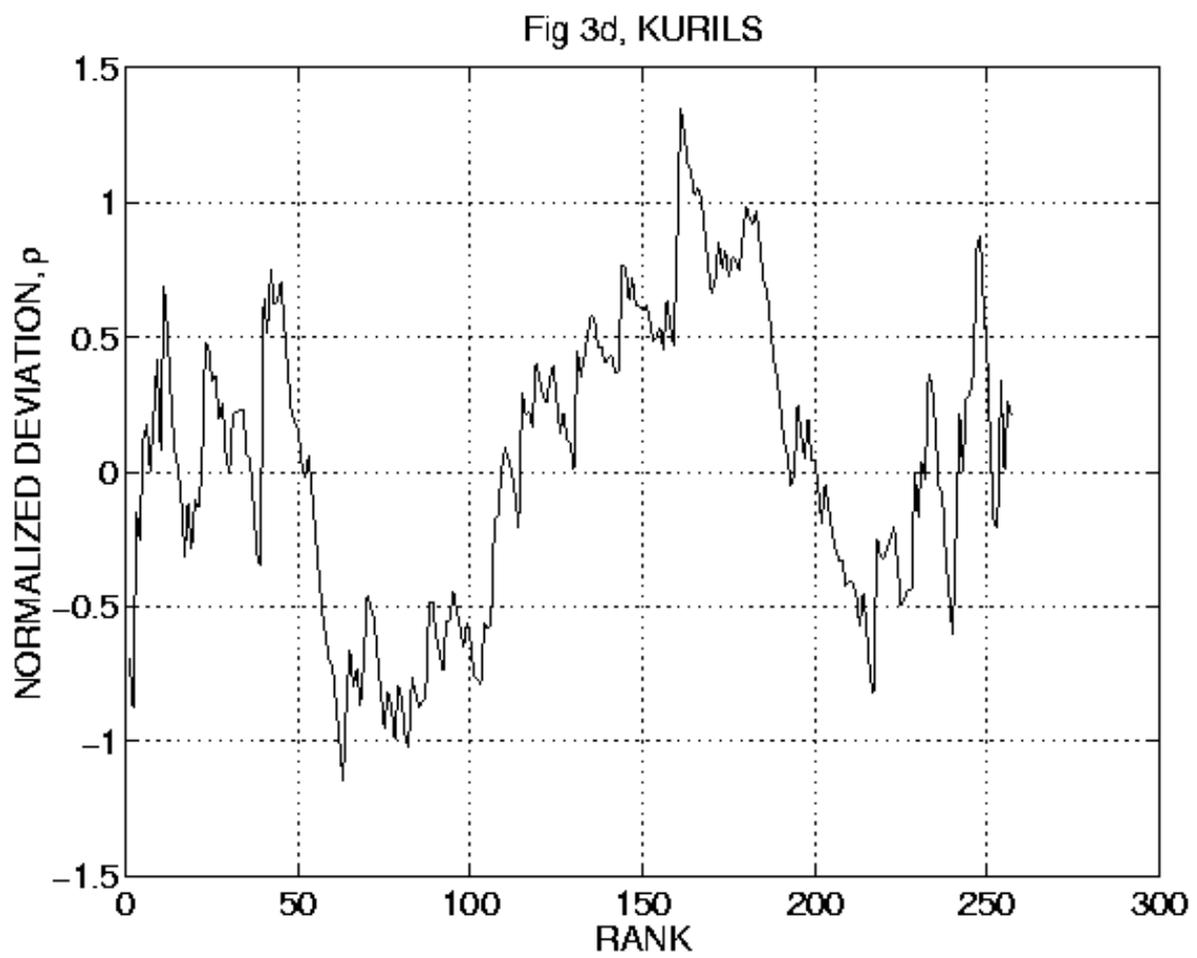

Fig 3d, KURILS

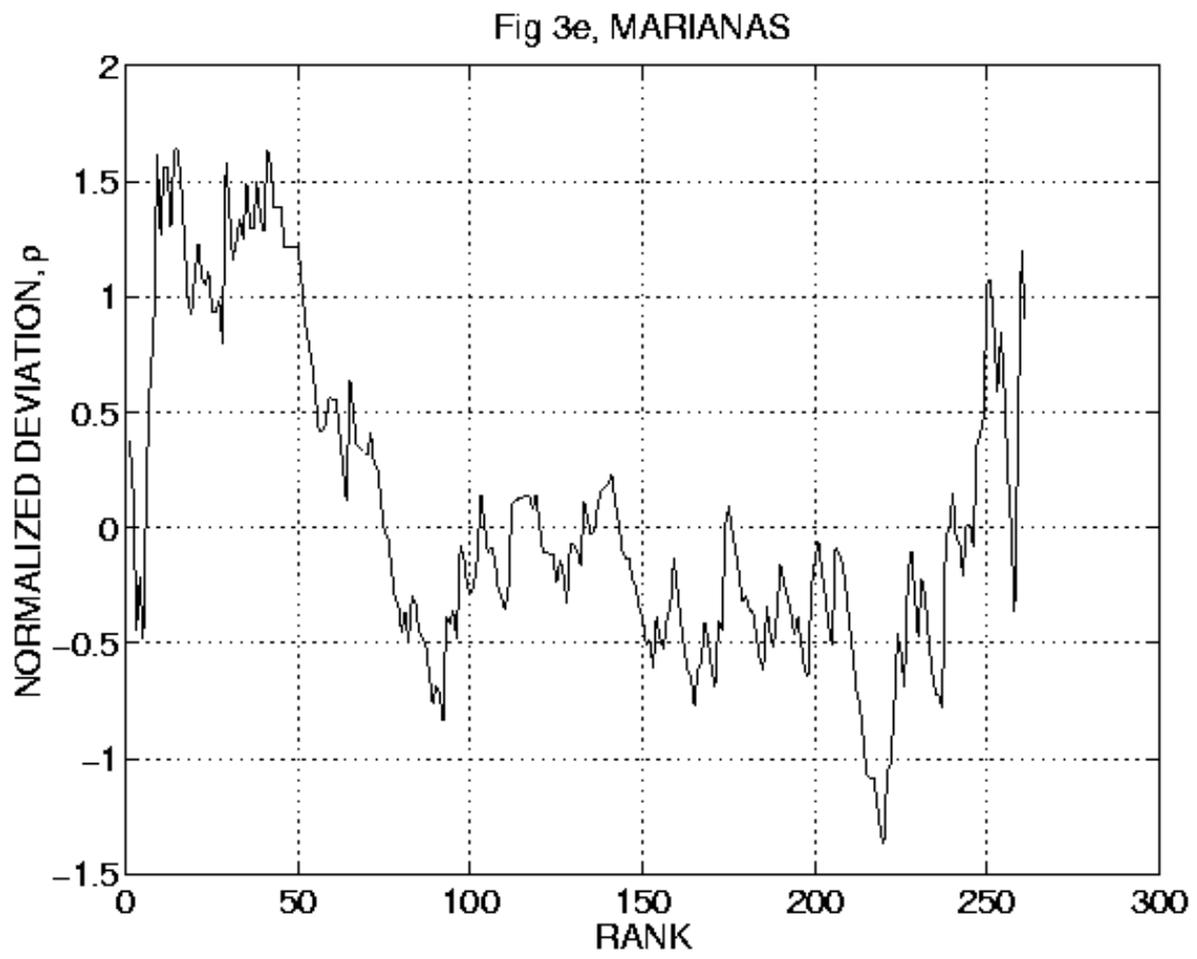

Fig 3e, MARIANAS

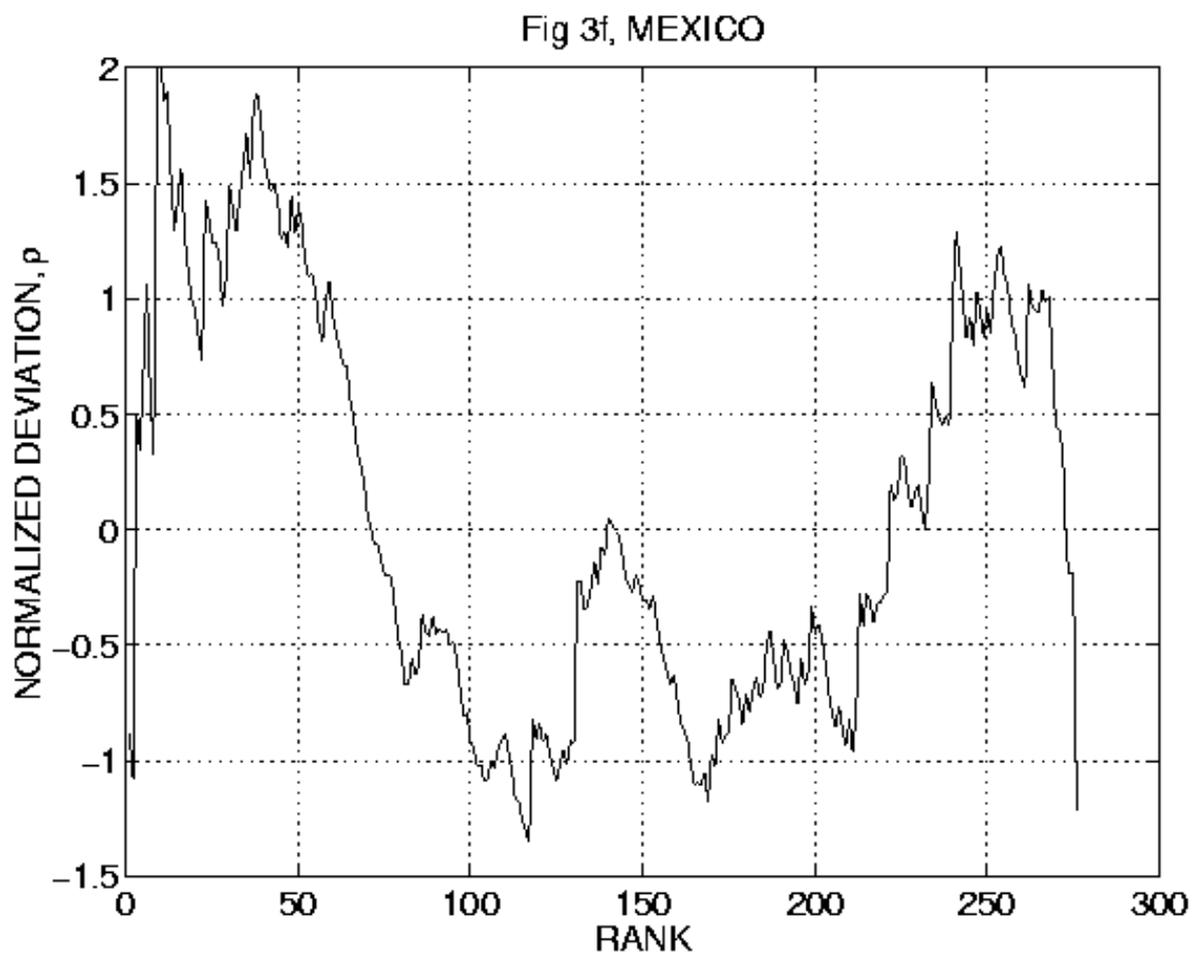

Fig 3f, MEXICO

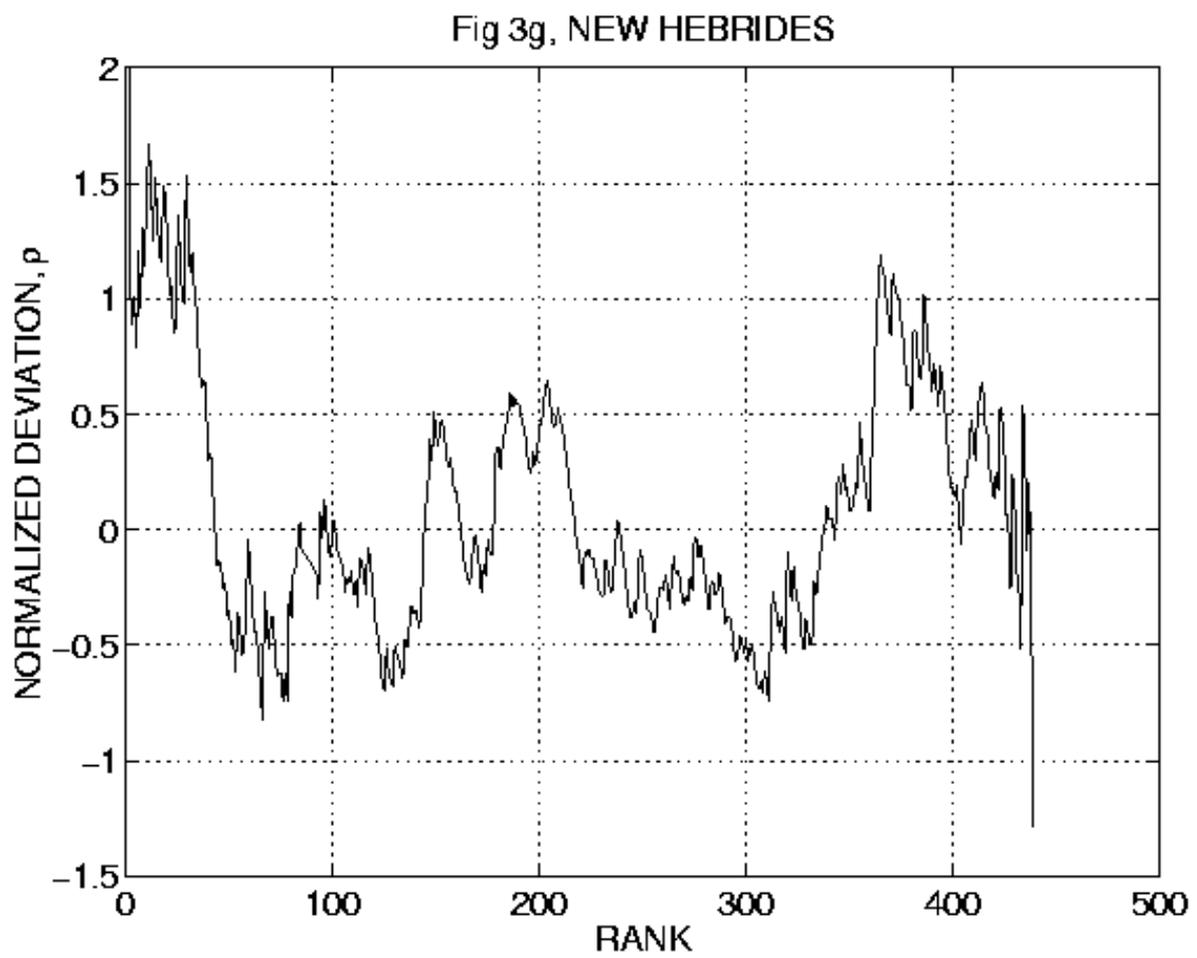

Fig 3g, NEW HEBRIDES

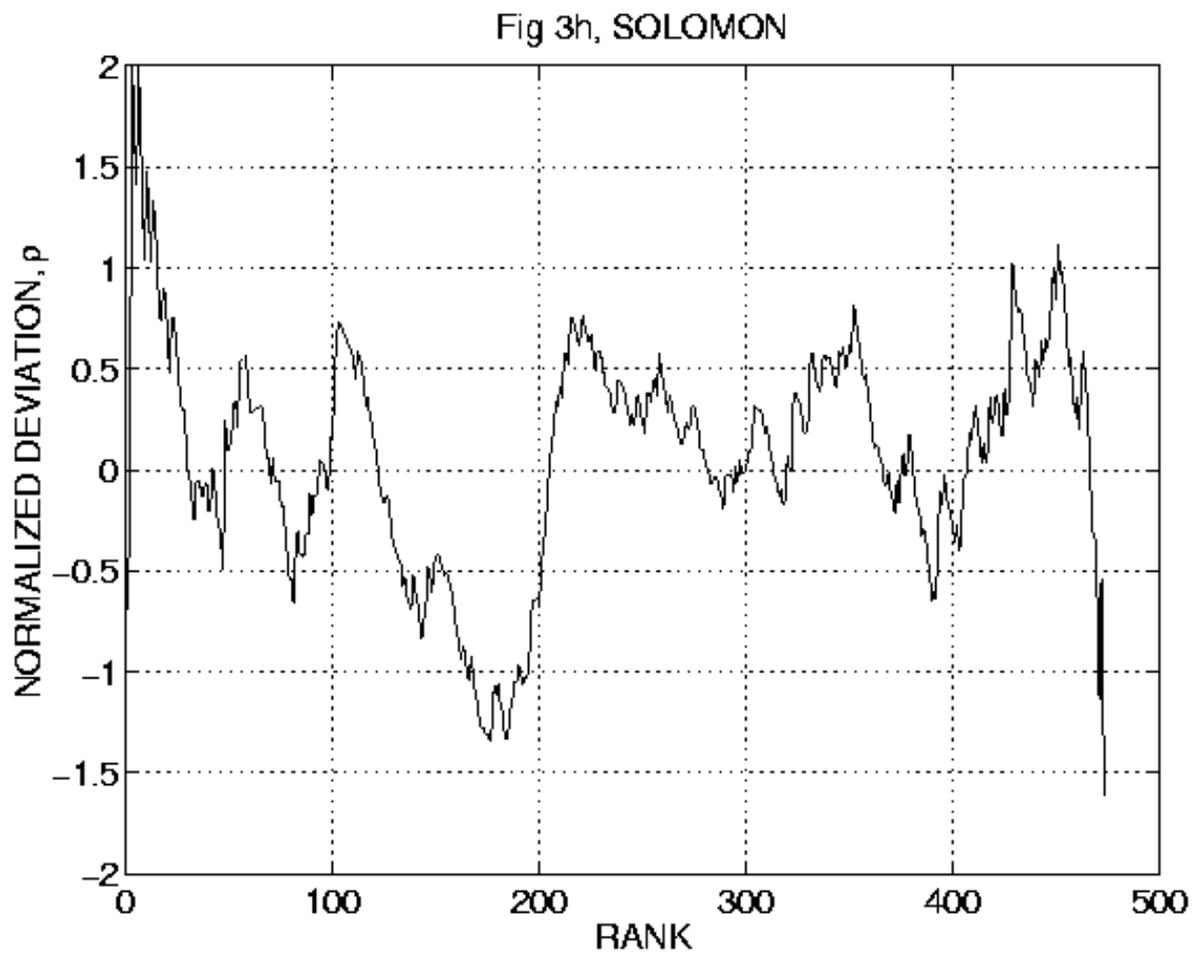

Fig 3h, SOLOMON

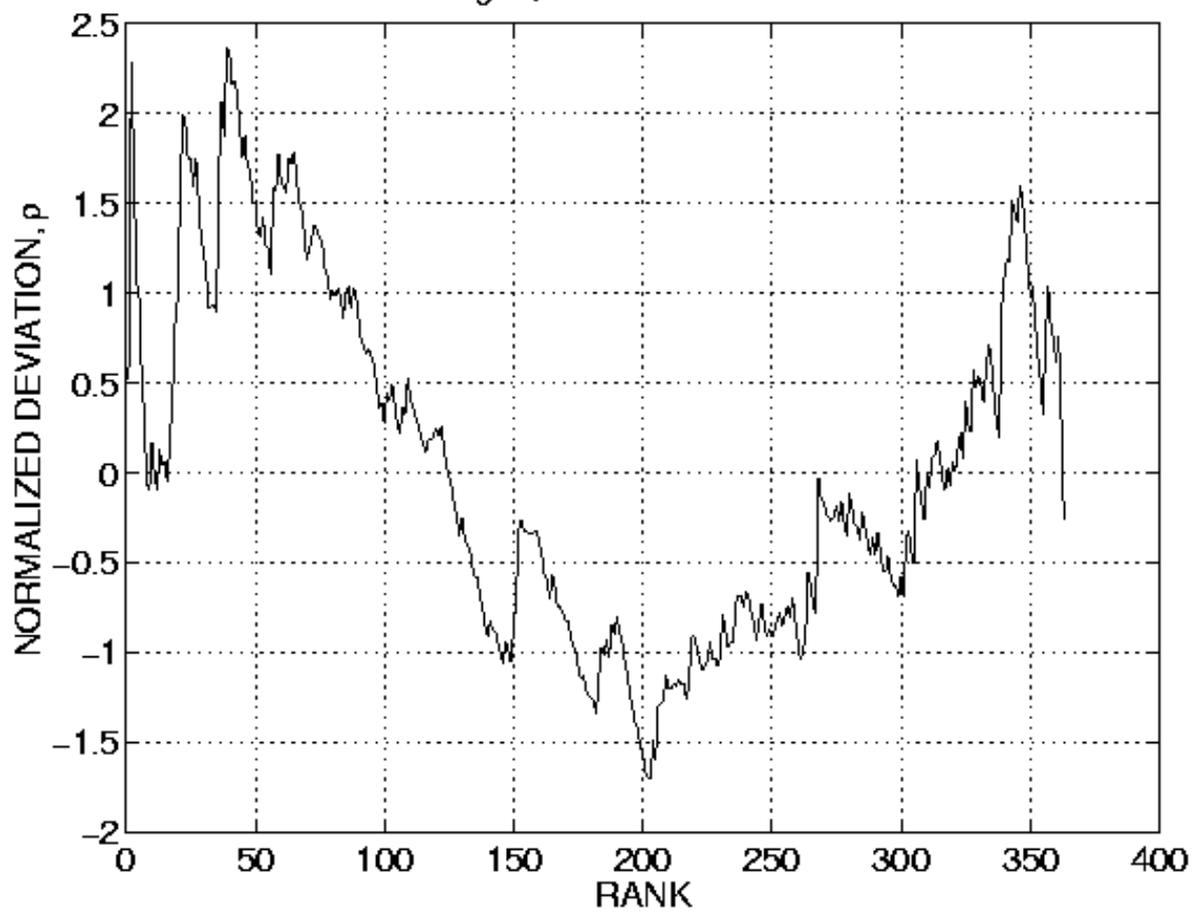

Fig 3i, SOUTH AMERICA

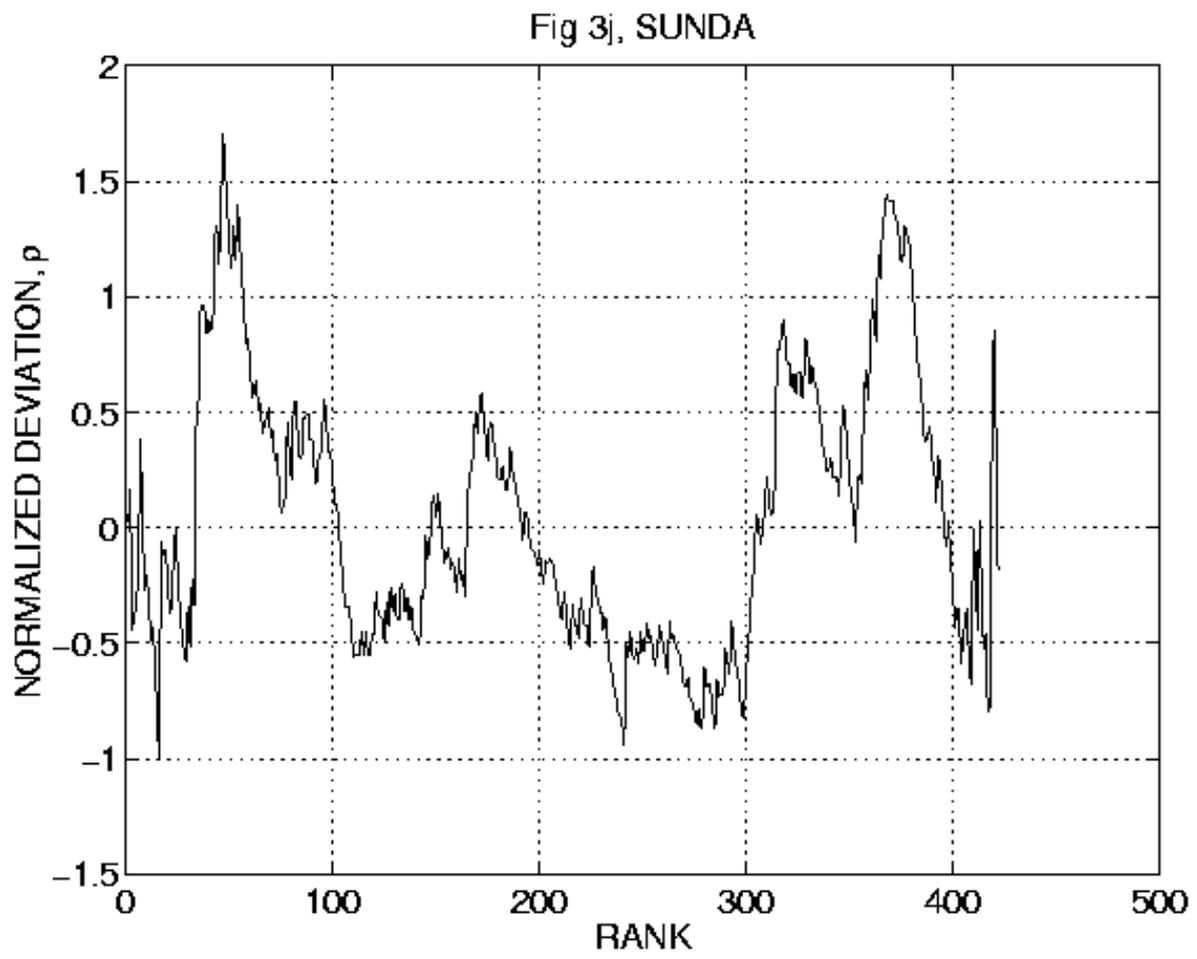

Fig 3j, SUNDA

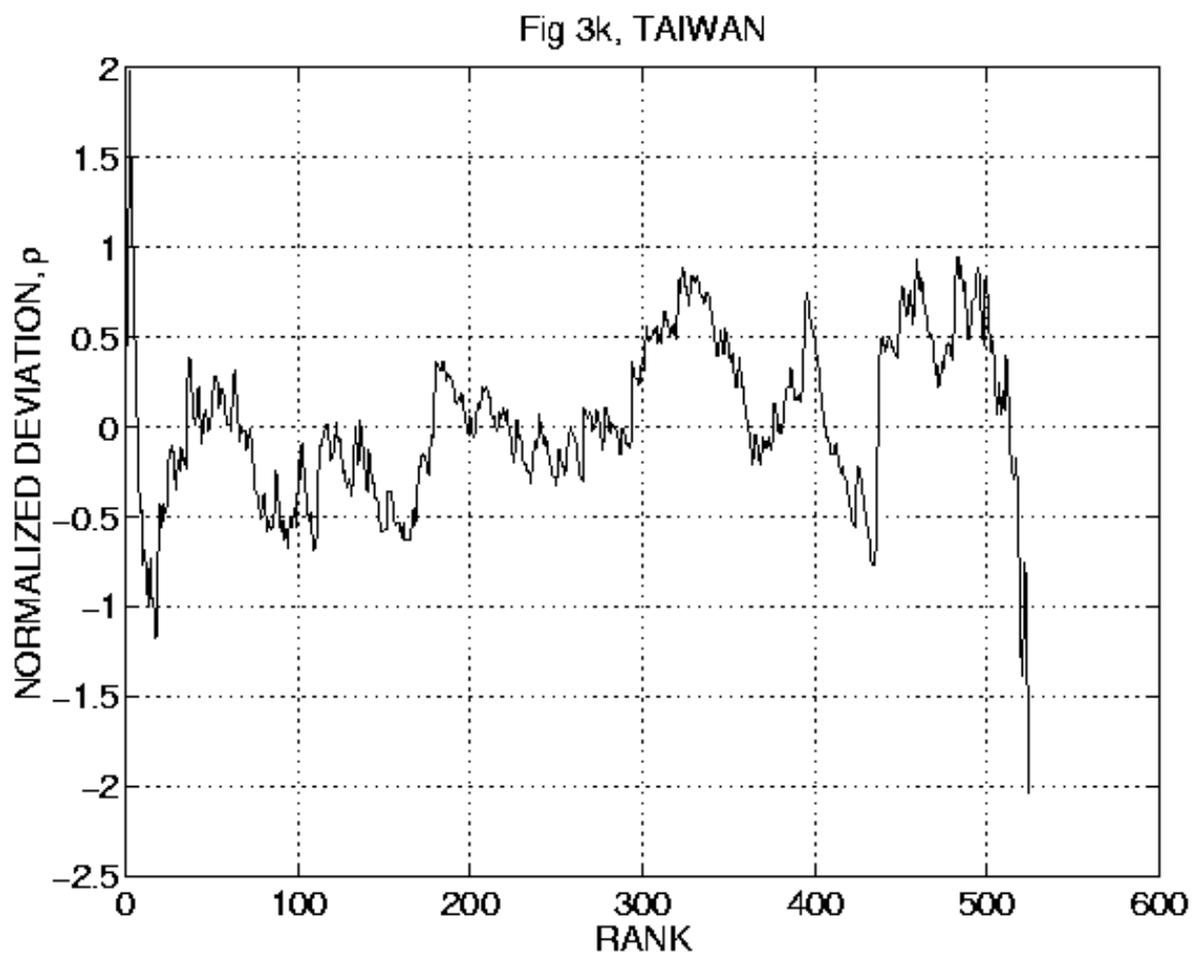

Fig 3k, TAIWAN

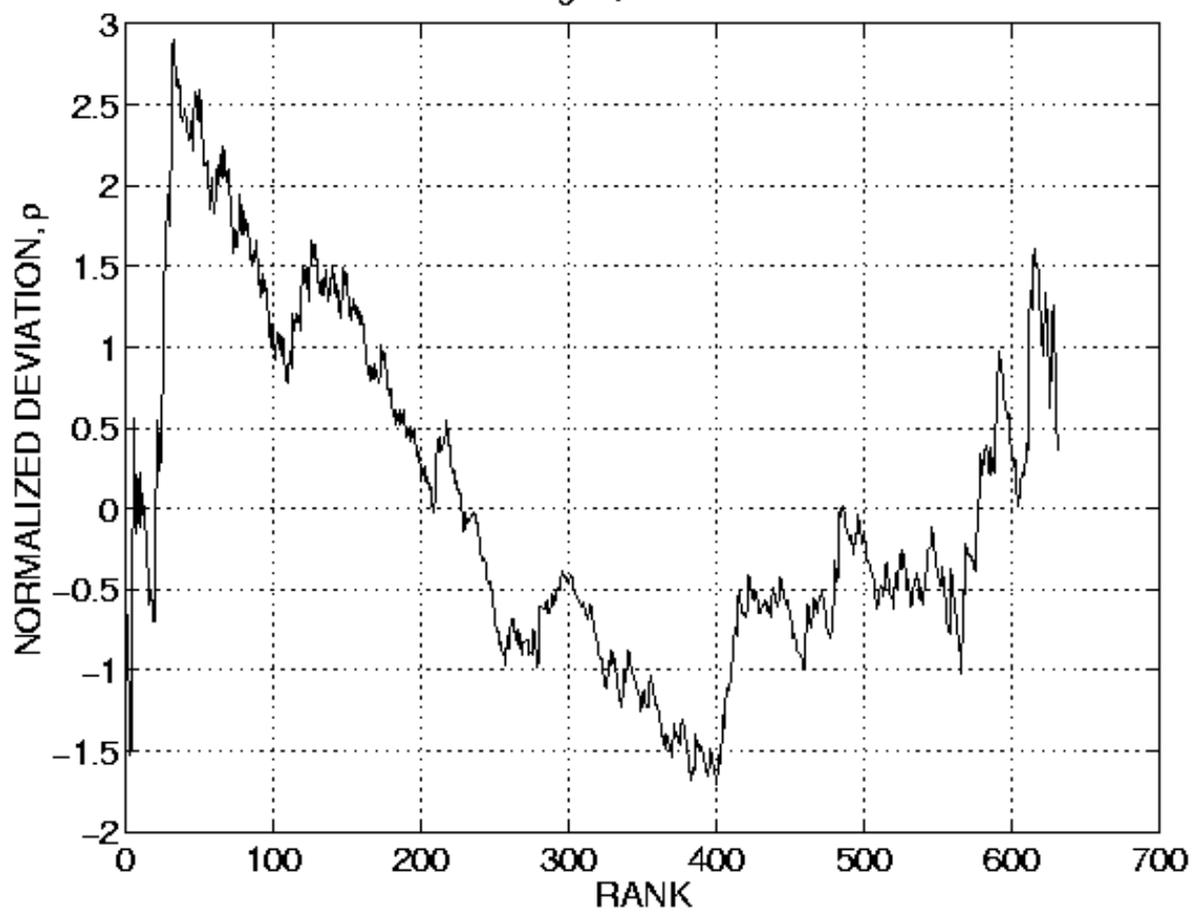

Fig 3l, TONGA

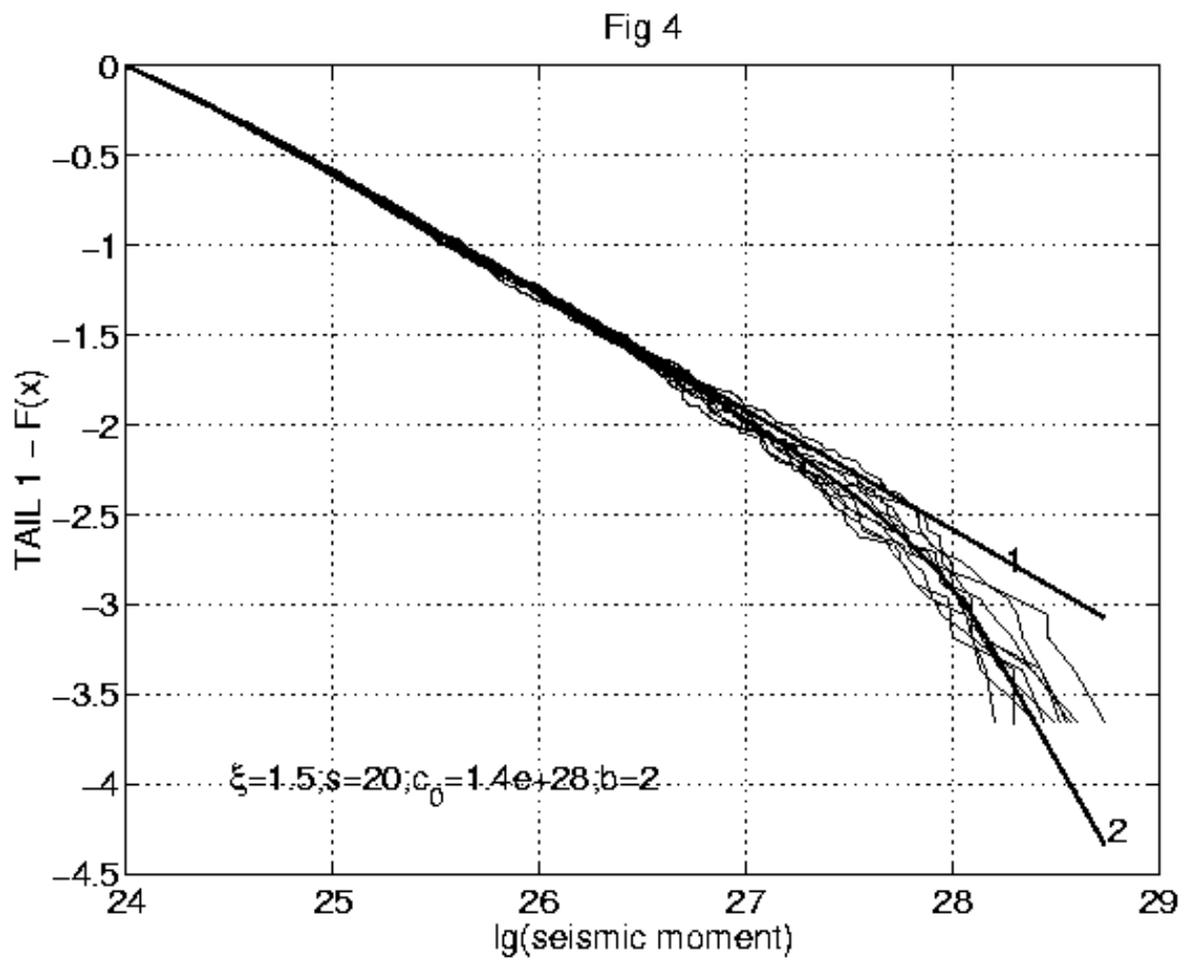

Fig 4

$\xi=1.5; s=20; c_0=1.4e+28; b=2$

Fig 5, lg C